\documentclass{astlb}
\usepackage{apjfonts}
\pdfoutput=1
\usepackage{amsmath}
\usepackage{graphicx}
\usepackage{natbib}
\usepackage{color}
\usepackage[hyperfootnotes=false]{hyperref}
\usepackage{multirow}
\definecolor{darkblue}{rgb}{0,0,0.9}

\newcommand{\Hefn}{\mbox{He\,II $\lambda4686$}}
\newcommand{\CIIIfn}{\mbox{C\,III\,/N\,III $\lambda4625$-$4658$}}
\newcommand{\HeIfn}{\mbox{He\,I $\lambda4922$}}

\newcommand{\solarmass}{\mbox{${\rm M_{\odot}}$}}

\newcommand{\He}{\mbox{He\,II}}
\newcommand{\CIII}{\mbox{C\,III\,/N\,III}}
\newcommand{\HeI}{\mbox{He\,I}}
\newcommand{\Hb}{\mbox{H$\beta$}}

\begin{document}
\journalinfo{2013}{39}{12}{826}[843] 
\title{\bf \Large Superbroad Component in Emission Lines of SS~433 }

\author{P.~S.~Medvedev \email{tomedvedev@iki.rssi.ru} \address{1}, S.~N.~Fabrika \address{2}, V.~V.~Vasiliev \address{3}, V.~P.~Goranskij \address{4} and E.~A.~Barsukova \address{2}
\addresstext{1}{Space Research Institute, ul. Profsoyuznaya 84/32, Moscow, 117997 Russia} 
\addresstext{2}{Special Astrophysical Observatory, Russian Academy of Sciences, Karachai-Cherkessian Republic,Nizhnii Arkhyz, 369167 Russia}
\addresstext{3}{Moscow State University, Moscow, 199992}
\addresstext{4}{Sternberg Astronomical Institute, Universitetskii pr. 13, Moscow, 119991 Russia}
}

\shortauthor{Medvedev et al.}
\shorttitle{SBC in Emission Lines of SS~433 }
\submitted{April 26, 2013}

\begin{abstract}
We have detected new components in stationary emission lines of SS 433; these are the superbroad components that are low-contrast substrates with a width of 2000--2500 km $\mbox{s}^{-1}$ in \HeIfn\ and \Hb\ and 4000--5000 km $\mbox{s}^{-1}$ in \Hefn. Based on 44 spectra taken during four years of observations from 2003 to 2007, we have found that these components in the \He\ and \HeI\ lines are eclipsed by the donor star; their behavior with precessional and orbital phases is regular and similar to the behavior of the optical brightness of SS 433. The same component in \Hb\ shows neither eclipses nor precessional variability. We conclude that the superbroad components in the helium and hydrogen lines are different in origin. Electron scattering is shown to reproduce well the superbroad component of \Hb\ at a gas temperature of 20--35 kK and an optical depth for Thomson scattering $\tau \approx$ 0.25--0.35. The superbroad components of the helium lines are probably formed in the wind from the supercritical accretion disk. We have computed a wind model based on the concept of Shakura-Sunyaev supercritical disk accretion. The main patterns of the \He\ line profiles are well reproduced in this model: not only the appearance of the superbroad component but also the evolution of the central two-component part of the profile of this line during its eclipse by the donor star can be explained.

\keywords{SS~433, close X-ray binaries, supercritical accretion, emission line formation.}

\end{abstract}

\section{INTRODUCTION}
\label{sec:intro}

SS~433 is the only known supercritical accretor in our Galaxy. This object is a massive eclipsing close binary system with an orbital period of 13.1 days \citep[for a review, see][]{fabrika04}. The donor star overfills its critical Roche lobe and transfers mass to the relativistic component (very likely a black hole) on the thermal time scale; the mass transfer rate from the donor to the accretion disk is $\dot M \sim 10^{-4}$\,\solarmass\, $\mbox{yr}^{-1}$.
 
The ultraluminous X-ray sources \citep{feng11} observed in external galaxies can be other examples of supercritical accretion disks. It is very likely that these objects are supercritical accretors like SS~433 \citep{fabrika01}, but their orientation is such that an observer can see the bottom of the supercritical disk funnel. Supercritical accretion is probably a necessary element for the growth of supermassive black holes \citep{volonteri05} at early stages of the increase in quasar mass. The supercritical regime can be of fundamental importance not only for the black hole growth efficiency but also for the feedback on the galaxies and the formation of galaxies and clusters of galaxies via jets and winds. Given the importance of these processes, detailed studies of SS~433 and the structure of gas flows in this system are needed, because no other bright and close examples of supercritical accretion disks have been found.

Despite the large number of studies devoted to SS 433, the mass of the relativistic star in this system has not been measured reliably; the spread in mass determinations for the compact object is from 2 to 15 \solarmass\ (see \citealt{fabrika90,hillwig04,blundell08,kubota10}; see, however, \citealt{goranskij11}). The system's luminosity has been measured much more reliably \citep{cherep02, fabrika08}, $L_{bol} \sim 10^{40}$\,erg $\mbox{s}^{-1}$ with its peak in the ultraviolet. Since the observed luminosity would be too high for a neutron star, a black hole with a supercritical accretion disk is believed to be in SS 433. For a black hole with a mass of $\sim 10$\,\solarmass, the rate of gas accretion into the disk of SS~433 roughly corresponds to 300--500 Eddington accretion rates.

The main energy release in SS~433 is accounted for by the relativistic component or, more precisely, its supercritical accretion disk. The observational manifestations of this system are completely determined by the disk orientation. Almost all of the accretion energy must be released in the hard X-ray range, but the X-ray luminosity of SS~433 ($L_x \sim 10^{36}$\,erg $\mbox{s}^{-1}$) is much lower than the bolometric one. The initial hard X-ray emission is thermalized in a powerful wind outlowing from the inner regions of the supercritical disk. The apparent size of the wind photosphere for an observer is $\sim 10^{12}$\,cm. The system's orientation is such that we cannot see the funnel base even at the times of best funnel visibility during the precessional motion of the disk. If the funnel bottom were seen, then SS~433 would probably be the most luminous X-ray source in the Galaxy with $L_x \geq L_{bol}$. An increase in the X-ray luminosity is predicted through the geometric collimation of emission by the supercritical disk funnel
      
SS 433 has precessing relativistic jets moving with a constant velocity $v_J \approx 0{.}26 c$ that are formed in the funnel of the supercritical accretion disk and closely follow the disk and funnel orientation. The so-called ``moving'' or ``relativistic'' hydrogen and \HeI\ lines are formed in the jets; these emission lines move over the spectrum in accordance with the orientation of the jets relative to the observer. The spectrum of SS~433 exhibits weak absorption lines from which the orbital motion of the donor star was measured \citep{gies02,hillwig04,cherep05,kubota10}. However, the extended gaseous envelope that outlows from the donor but is no longer gravitationally bound to it may make a noticeable contribution in the absorption line. Even at the times of the deepest (``total'') eclipses of the accretion disk by the donor, the contribution from the supercritical disk to the total brightness of the system is larger than the contribution from the donor \citep{gies02,hillwig08,goranskij11}. According to this contribution, the extended envelope of the donor will distort signiicantly the donor's absorption lines.

The brightest emission lines in SS~433 are the hydrogen ones that originate in the wind outflowing from the accretion disk and in the gas lost by the system through the Lagrangian point behind the disk \citep{blundell08}. The \HeI\ and Fe\,II emission
lines originate in the same medium. All these lines show orbital motion with different amplitudes and with a phase lag relative to the instantaneous position of the accretion disk \citep{crampton81,kopylov89}. For this reason, it is difficult to measure the system's mass using these lines.

The only line that reflects the motion of the relativistic component (in fact, the wind that is formed in the disk) is the \He\ line; it was used to measure the system's mass function \citep{crampton81,fabrika90}. The width of the \He\ line is \mbox{$\mbox{FWHM} = 600$--$1000$}\,km $\mbox{s}^{-1}$; obviously, it is formed in the hottest part of the wind, possibly closer to the disk axis. The orbital radial velocity curve depends on the precessional phase, i.e., on the disk inclination to the line of sight. Knowing the behavior of this line is very important for understanding the reliability of measuring the masses.

Here, we investigate the He\,II line profile based on
our and archival best-quality spectra and find a new
component in the profile of this line that we call a
``superbroad'' component (SBC). We find the same
components in the He I and hydrogen lines. We
investigate the behavior of these components using
the H$\beta$ and He\,I\,$\lambda 4921$ lines as an example. In the
final part of the paper, we discuss the interpretation of
the SBCs in these lines.

\section{OBSERVATIONAL DATA}
\label{s_obs}

Our observational material consist of 44 optical
spectra for SS~433 taken with different telescopes
from 2003 to 2007. All spectra are only of good
quality; these are the total spectra obtained during a
single night. The dates of our spectroscopic observations,
the precession and orbital phases, and the
telescopes/instruments are given in Table~\ref{tab:1}.

In 2003, observations were carried out at the
6-m BTA telescope with the UAGS spectrograph \citep{cherep05}; the spectral resolution
was 4~\AA. At the same telescope but with the SCORPIO
spectrograph \citep{afanasiev05},
spectra were taken in 2004--2007; the spectral resolution
was 2 (2004), 2.5--3 (2005, 2006), and $\approx 5$~\AA\ 
(2007). In October 2007, simultaneously with BTA,
spectroscopy for SS~433 was performed at the Subaru
telescope with the FOCAS spectrograph \citep{kubota10}; the spectral resolution was 1.5~\AA.

Our spectra were supplemented with archival
data. These are the 2004 spectra from the 4.2-m
WHT telescope taken with the ISIS spectrograph
\citep{clegg81} with a resolution of 0.7~\AA. Here, we
selected a continuous set of observations consisting
of six nights that covers the precessional phases
when the line of sight was in the disk plane. We
also used the archival data obtained in 2006 at
the 8-m Gemini-North telescope with the GMOS
spectrograph \citep{hillwig08}; the spectral
resolution in these data was $\approx 1$~\AA. The difference in
the spectral resolution of our data does not affect the
identification and analysis of SBCs in the line profiles.

\begin{table*}
  \renewcommand{\arraystretch}{1.2}
\centering
\caption{Log of spectroscopic observations for SS~433. The columns present the dates of observations, the orbital and precessional phases, the $B$ magnitudes, the photometric
coefficients, and the telescope/intrument.}
\label{tab:1}
 \medskip
 \small
\begin{tabular}{ccccccl}
\hline
 Date   &JD 2450000+&$\phi$& $\psi$& $B$ & $f$ & Telescope/spectrograph\\   
\hline
May 9, 2003& 2769.51 &0.885 & 0.079& 16.25 & 0.796& BTA/UAGS\\
May 10, 2003& 2770.50 &0.961 & 0.085& 16.97 & 0.411&BTA/UAGS\\
May 11,2003& 2771.49 &0.036 & 0.091& 16.86 & 0.454& BTA/UAGS \\
May 12, 2003& 2772.51 &0.114 & 0.097& 16.44 & 0.667&BTA/UAGS \\
May 13, 2003& 2773.49 &0.189 & 0.103& 15.90 & 1.096&BTA/UAGS\\
June 29, 200& 3186.66 &0.772 & 0.649& 16.73 & 0.510&WHT/ISIS\\
June 30,2004& 3187.56 &0.841 & 0.655& 16.73 & 0.510&WHT/ISIS\\
July 1,2004& 3188.65 &0.924 & 0.662& 16.78 & 0.490& WHT/ISIS\\
July 2, 2004& 3189.67 &0.002 & 0.668& 16.87 & 0.450&WHT/ISIS\\
July 3, 2004& 3190.49 &0.065 & 0.673& 16.93 & 0.425&WHT/ISIS\\
July 4, 2004& 3191.55 &0.146 & 0.680& 16.80 & 0.480&WHT/ISIS\\
Aug. 22, 200& 3240.33 &0.874 & 0.980& 16.33 & 0.735&BTA/SCORPIO\\
Aug. 23, 200& 3241.32 &0.950 & 0.986& 16.96 & 0.412&BTA/SCORPIO\\
Sep. 7, 2004& 3256.29 &0.094 & 0.078& 16.31 & 0.750&BTA/SCORPIO \\
Sep. 8, 2004& 3257.27 &0.169 & 0.084& 16.21 & 0.824&BTA/SCORPIO\\
Sep. 9, 2004& 3258.29 &0.247 & 0.091& 16.09 & 0.922&BTA/SCORPIO\\
June 7,2005& 3529.47 &0.976 & 0.762& 17.17 & 0.339& BTA/SCORPIO\\  
June 8,2005& 3530.45 &0.051 & 0.768& 16.92 & 0.427& BTA/SCORPIO\\     
June 9,2005& 3531.44 &0.127 & 0.774& 16.83 & 0.466& BTA/SCORPIO\\
June 10,2005& 3532.47 &0.205 & 0.780& 16.60 & 0.576&BTA/SCORPIO\\
May 20, 2006& 3876.50 &0.503 & 0.900& 16.57 & 0.590&BTA/SCORPIO  \\
May 21,2006& 3877.44 &0.575 & 0.906& 16.31 & 0.753& BTA/SCORPIO\\
May 22, 2006& 3878.49 &0.655 & 0.913& 16.24 & 0.801&BTA/SCORPIO \\
May 23, 2006& 3879.48 &0.731 & 0.919& 16.05 & 0.956&BTA/SCORPIO \\
May 24, 2006& 3880.44 &0.804 & 0.925& 16.04 & 0.966&BTA/SCORPIO\\
May 25, 2006& 3881.40 &0.877 & 0.931& 16.13 & 0.889&BTA/SCORPIO\\
May 26, 2006& 3882.40 &0.954 & 0.937& 16.51 & 0.625&BTA/SCORPIO  \\ 
May 30, 2006& 3886.49 &0.266 & 0.962& 16.28 & 0.774&BTA/SCORPIO \\
May 31, 2006& 3887.49 &0.343 & 0.968& 16.29 & 0.767&BTA/SCORPIO\\
June 7,2006& 3893.99 &0.840 & 0.008& 16.18 & 0.847& Gemini/GMOS\\
June 8,2006& 3895.04 &0.920 & 0.015& 16.44 & 0.667& Gemini/GMOS\\
June 9,2006& 3895.96 &0.990 & 0.020& 16.70 & 0.525& Gemini/GMOS\\
June 10,2006& 3897.02 &0.071 & 0.027& 16.74 & 0.506&Gemini/GMOS\\
June 11,2006& 3898.01 &0.147 & 0.033& 16.39 & 0.698&Gemini/GMOS \\
June 12,2006& 3899.03 &0.225 & 0.039& 16.29 & 0.766&Gemini/GMOS\\
June 13,2006& 3900.03 &0.301 & 0.045& 16.05 & 0.955&Gemini/GMOS\\
Oct. 4, 2007& 4378.23 &0.855 & 0.992& 16.10 & 0.914&BTA/SCORPIO \\
Oct. 5, 2007& 4379.23 &0.931 & 0.998& 16.43 & 0.672&BTA/SCORPIO\\
Oct. 6, 2007& 4379.76 &0.972 & 0.002& 16.74 & 0.510&Subaru/FOCAS \\
Oct. 6, 2007& 4380.20 &0.005 & 0.004& 16.93 & 0.420&BTA/SCORPIO\\
Oct. 7, 2007& 4380.78 &0.050 & 0.008& 16.80 & 0.480&Subaru/FOCAS\\
Oct. 7, 2007& 4381.22 &0.083 & 0.011& 16.44 & 0.665&BTA/SCORPIO\\ 
Oct. 8, 2007& 4381.84 &0.131 & 0.014& 16.21 & 0.820&Subaru/FOCAS \\
Oct. 10,2007& 4383.77 &0.278 & 0.026& 16.03 & 0.970&Subaru/FOCAS \\
\hline
\end{tabular}

\end{table*}

All data were reduced in a standard way, including
the bias subtraction, flat fielding, the wavelength
calibration of spectra, and the extraction of one-dimensional
spectra. We performed all of the reduction
operations in the \mbox{ESO MIDAS} system\footnote{http://www.eso.org/sci/software/esomidas/} \citep{warmels92}. Since we study only the line profiles
here, all spectra were normalized to the continuum
level. The continuum was always determined in the
same way and constructed from the same reference
points in the spectrum.

To calculate the orbital and precessional phases,
we used the ephemerides from \cite{goranskij11}: 
\[\mbox{Min\,I} = \mbox{{\rm JD\,}} 2450023.746 +13^d.08223 \]
\[ \mbox{{\rm T}}_{\mbox{{\scriptsize max}}} = \mbox{{\rm JD\,}} 2449998.0+162^d.278\] 

Here, the orbital phase $\phi = 0$ (Min\,I) corresponds
to the mid-eclipse of the accretion disk (relativistic
component) by the donor star, and the precessional
phase of the disk $\psi = 0$ (${\rm T_{max}}$ or ``${\rm T_3}$ moment'')
corresponds to the maximum disk opening toward the
observer or the maximum separation of the relativistic
lines in the spectrum. The moments when the relativistic
jets of SS~433 lie in the plane of the sky are called
crossovers; they correspond to precessional phases
$\psi = 0.33$ and 0.66 (``edge-on disk'').

Most of the observations were carried out near
precessional phases of about zero, because the gas
outflowing from the accretion disk does not cover
the donor only at these phases. These are the most
optimal phases for investigating the orbital motion of
the donor. Nevertheless, some of the observations
in 2004 and the observations in 2005 were carried out
when the disk was seen edge-on. During the 2006
observations at both BTA and Gemini, SS~433 was in
an active state. Very powerful sporadic gas ejections
from the disk occur in its active state \citep{fabrika03,fabrika04}; the object is very
difficult to investigate. The remaining observations
were carried out in the passive state of SS~433.

The photometry was obtained on the same night
as the spectroscopy and was offset in time from the
spectroscopy by no more than 2--3 h. The object's
sporadic variability in such time interval does not exceed 10\,\%.
Usually, we used the images obtained immediately
before the spectroscopy. Table~1 gives the $B$ magnitude
measurements for SS~433. The BTA spectroscopic
observations were accompanied by multicolor
photometric UBVRI observations with a 1-m
Zeiss telescope at the Special Astrophysical Observatory
of the Russian Academy of Sciences \citep{kubota10} or BVR observations with a 60-cm Zeiss
telescope at the Crimean Station of the Sternberg Astronomical Institute of the Moscow State University \citep{cherep05}. We also used
the direct images obtained during the spectroscopic
observations when the telescope was pointed toward
the object. The photometric data for the Gemini
observations were taken from \cite{kubota10};
these magnitudes were obtained by calibration from
the Sloan g$^{\prime}$ band to the $B$ band.

Table~1 gives the photometric coefficient $f$ calculated
from the formula $f = 2{.}512^{16.0 - B}$. This coefficient
allows for the variability of SS~433 in continuum
and is tied to $\mbox{B} = 16.0$, which corresponds to the
mean out-of-eclipse brightness of SS~433 at the precessional
phase of the largest disk inclination to the
observer. We worked only with the normalized spectra.
Brightness variations lead to variations in line
intensities and equivalent widths in the normalized
spectra. Lest the line intensities be distorted during
the brightness variations in SS~433, we used this
coefficient (unity was subtracted from the normalized
spectra, and they were multiplied by $f$). Accordingly,
all of the spectra here were corrected for the photometric
coefficient. In our case (and everywhere in this
paper), the relative line intensity or equivalent width
is the same, with their variability reflecting the actual
variability of the line fluxes.

Here, we study the \Hefn\ and \HeIfn\
emission lines, the Bowen \CIIIfn\ ($\lambda_{eff} = 4644$\,\AA) blend, and H$\beta$. Below, unless stated otherwise, the first three lines are called \He, \HeI, and \CIII, respectively. All these lines are in
the $B$ band. We disregard the variable contribution
from these emission lines to the B band, because it
is small. The strongest \Hb\ line has an equivalent
width of 40--50~\AA; the lines in the spectrum of SS~433
change only slightly and, therefore, the intrinsic variability
of these lines has virtually no effect on the total
flux in the $B$ band.

\section{SUPERBROAD COMPONENTS OF PROFILES}
\subsection{Analysis of Line Profiles}
\label{sec_an}

\begin{figure*}[t]
\centering
\includegraphics[width=0.6\textwidth]{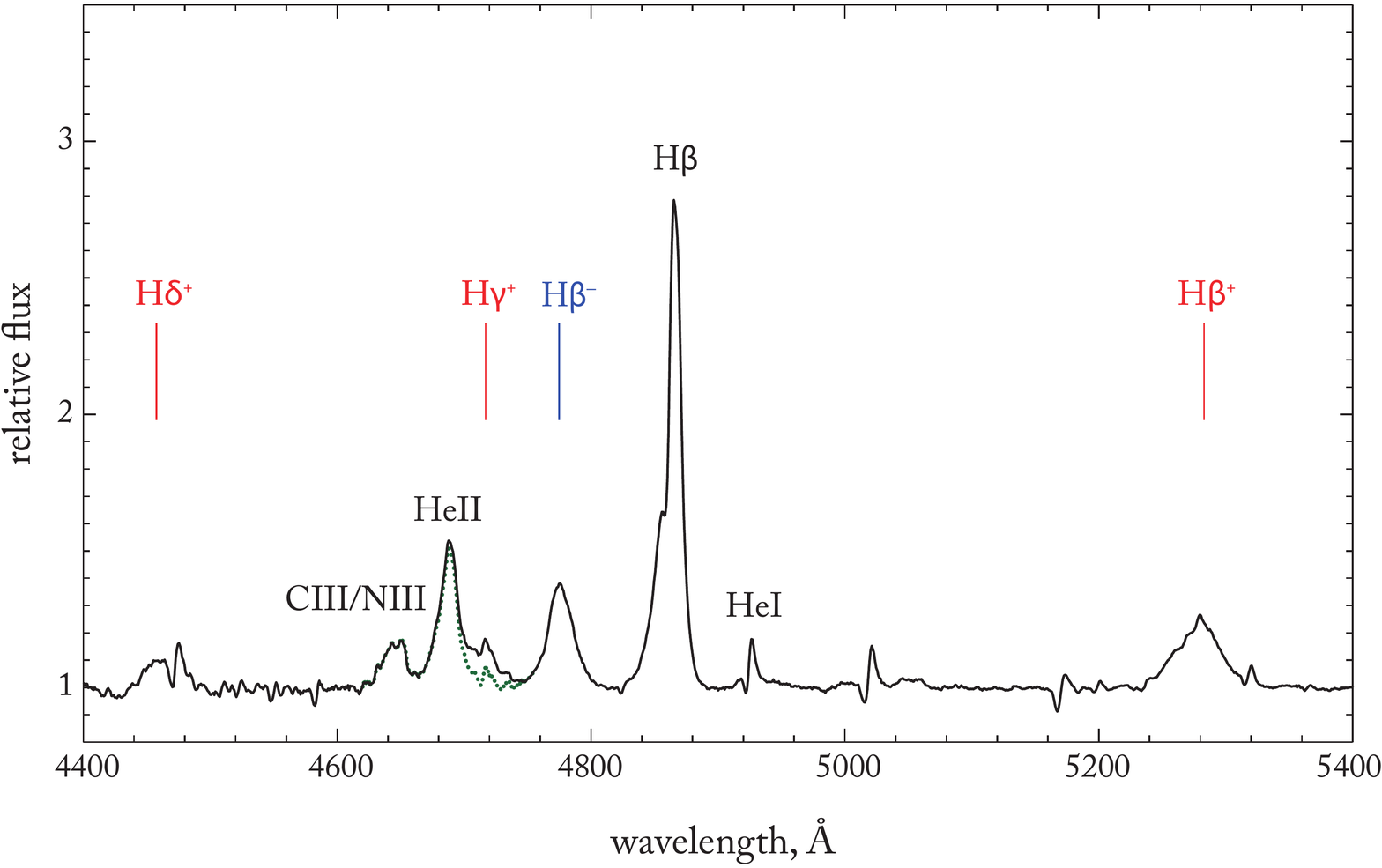}
\caption{BTA spectrum of SS~433 with ``moving'' and stationary lines taken on June 7, 2005. The green dotted line indicates the result of subtracting the relativistic H$\gamma^+$ emission line.}
\label{fig:RE}
\end{figure*}

Figure~\ref{fig:RE} presents a BTA optical spectrum of
SS~433. In addition to the bright stationary \Hb\ and
\He\ lines and the Bowen \CIII\ blend, the
spectrum exhibits lines from the relativistic jets that
manifest themselves as ``moving'' hydrogen and \HeI\
emission lines and move over the spectrum because of
a change in the jet inclination to the line of sight. Two
groups of relativistic emission lines that are formed
in the receding and approaching jets (in Fig.~\ref{fig:RE}, H$\beta^+$, H$\gamma^+$, H$\delta^+$, and H$\beta^-$, respectively) are observed. The location of the jets in space and the line positions in the spectrum are described by the kinematic model \citep{abell79,eikenberry01}.

When the relativistic emission lines blended with
the lines of interest to us (\He, \HeI\, and \Hb), we
subtracted the relativistic line profiles. For example,
the relativistic H$\gamma^+$ emission line in Fig.~\ref{fig:RE} is superimposed on the wing of the stationary \He\ line.
To subtract the relativistic emission line, we found
an unblended relativistic line in the same spectrum;
this is H$\beta^+$ in Fig.~\ref{fig:RE}. According to the position
and profile of H$\beta^+$, we determined the position of
H$\gamma^+$ blending the \He\ line. If there were several
relativistic lines in the spectrum for a given night,
then we determined the Balmer decrement and found
the intensity of the required relativistic line by interpolation.
If there was only one free relativistic emission
line in the spectrum, then the Balmer decrement for a
given precessional phase was taken as the mean from \cite{panferov97}. Figure~\ref{fig:RE} shows the corrected
spectrum with the removed relativistic emission line.

When investigating the He II line profile, we identified
the broad substrate of this line. The line profile
often looks a multicomponent one; as a rule, two
``narrow'' (FWHM $\sim 300$\,km $\mbox{s}^{-1}$) and two ``broad''
(FWHM $\sim 850$\,km $\mbox{s}^{-1}$) components are observed.
The new component is a very broad low-contrast
substrate with a width of about 60\,\AA\ or 4000 km $\mbox{s}^{-1}$,
which, nevertheless, is detected with confidence
(Fig. 2--4). We called it a ``superbroad'' component
(SBC). Similar components were detected in the \HeI\
and \Hb\ lines under study and in the \CIII\
blend, but the brightest SBC is observed in the
\He\ line. Note that the same component has been observed and already identified in H\,$\alpha$ \citep{dopita81, bowler10}.

Since the left wing of the \He\ line and the right
wing of the \CIII\ blend overlap, we cannot
reliably identify the SBCs of these lines separately.
We suggest that the SBC is a result of the addition of
the \He\ and \CIII\ SBCs. The centroid of this
combined SBC is approximately halfway between the
\He\ and \CIII\ lines (slightly shifted toward the
helium line).

We used two Gaussians to fit the SBCs of the
\He\ and \CIII\ blend lines. We selected the
reference points on the wings of the superbroad profile
blueward of the \CIII\ blend and redward of the
\He\ line (given the \HeI\ $\lambda4713$ line) as well as in the
gap between these lines. In all our spectra, the reference
points were always selected in the same places
of the profile. Since the parameters of each individual
Gaussian (\CIII\ and \He) cannot be accurate,
below we used only the sum of the intensities of both
SBCs. The \HeI\ and \Hb\ lines are single ones; the
SBCs in them are easier to analyze. In these lines,
the reference points were selected on the wings of the
superbroad profile.

\begin{figure*}
\centering
\includegraphics[width=0.65\textwidth]{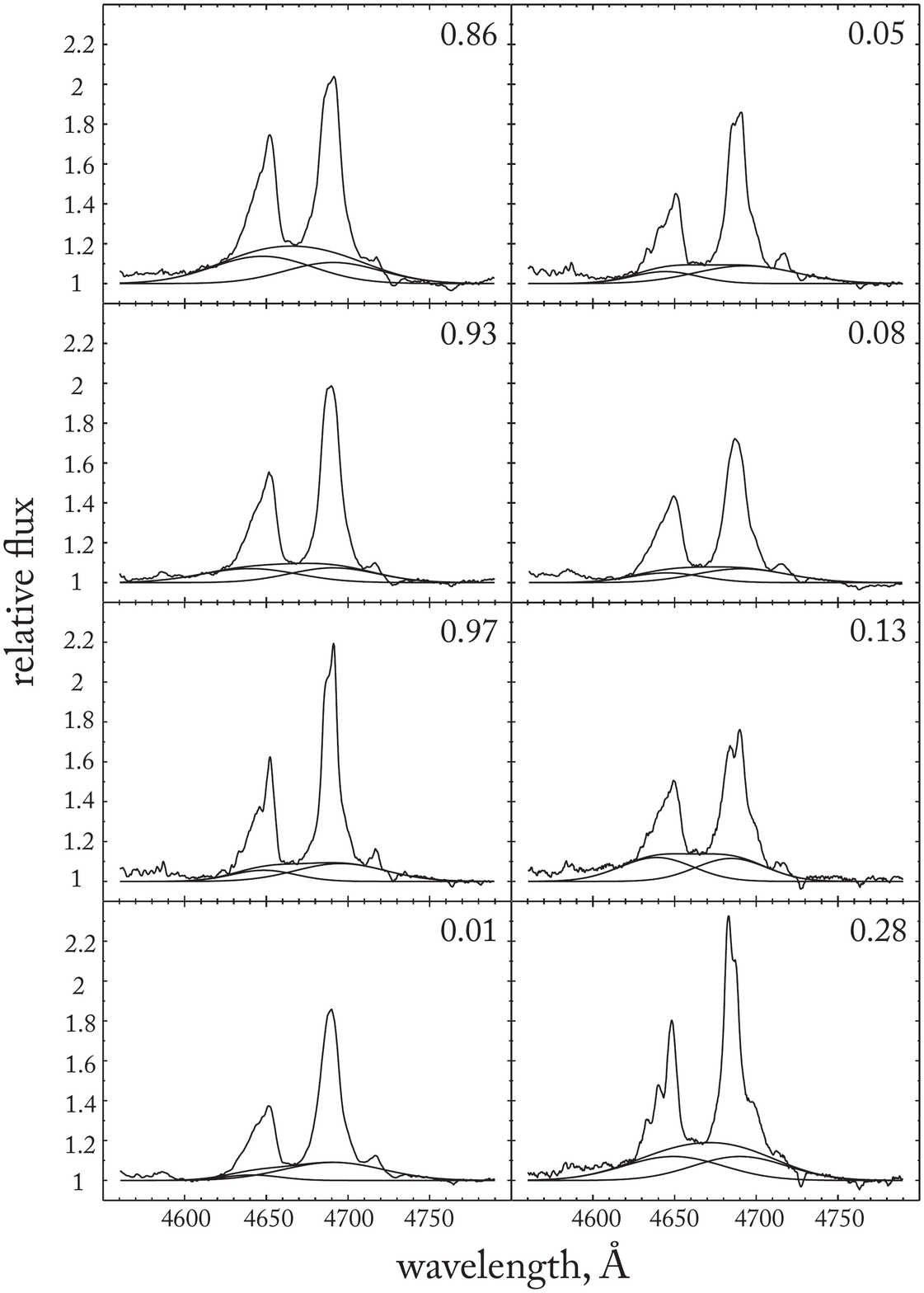}
\caption{\He ~and \CIII\ line profiles with the identified SBCs (two Gaussians and their sum) from the 2007 BTA and
Subaru spectra. The orbital phases are shown in the spectra. In the BTA spectra (phases 0.86, 0.93, 0.01, 0.08), the spectral
resolution is appreciably lower than that in the Subaru spectra.}
\label{fig:HeII_07}
\end{figure*}

\begin{figure*}
\centering
\includegraphics[width=0.65\textwidth]{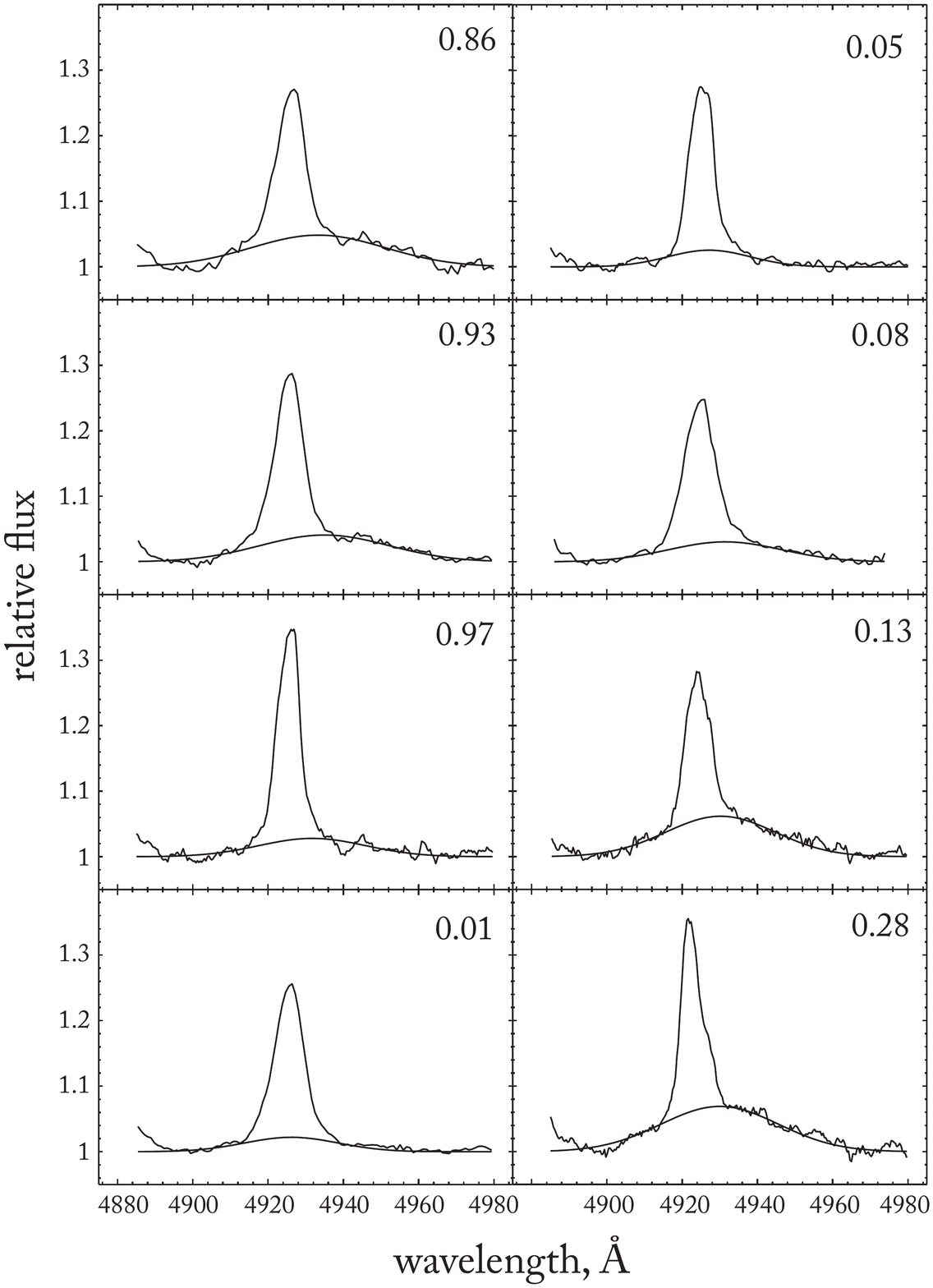}
\caption{\HeI\ line profiles with the identified SBCs (Gaussian). The rest is the same as in Fig.~\ref{fig:HeII_07}.}
\label{fig:HeI_07}
\end{figure*}

\begin{figure*}
\centering
\includegraphics[width=0.65\textwidth]{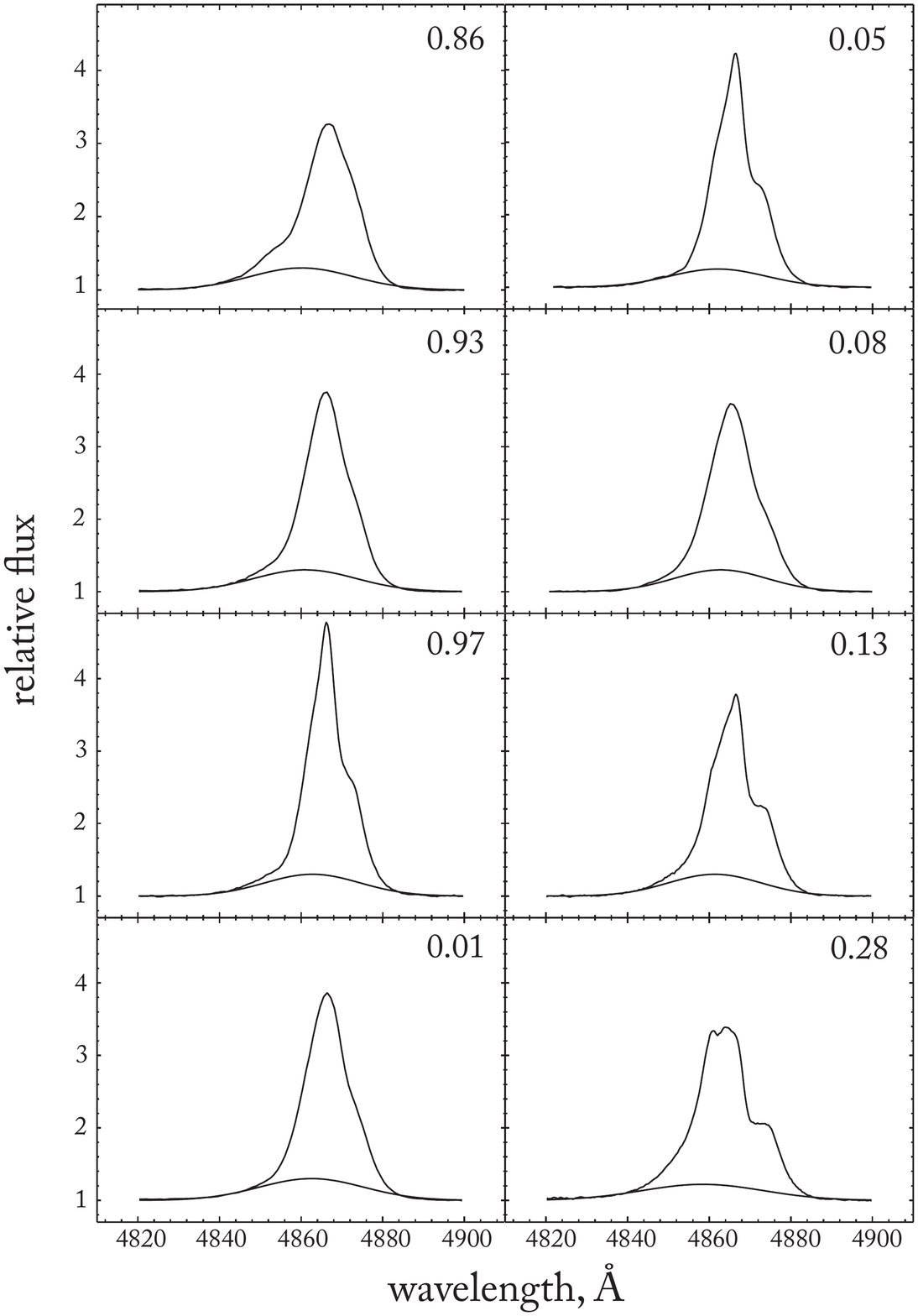}
\caption{\Hb\ profiles with the identified SBCs (Gaussian). The rest is the same as in Fig.~\ref{fig:HeII_07}.}
\label{fig:Hb_07}
\end{figure*}

In all our spectra, the \He\ line clearly exhibits
a SBC; the 2003 set of observations, in which the
SBC is weak, constitutes an exception. Figures~\ref{fig:HeII_07}--\ref{fig:Hb_07}
 present the spectra of SS 433 with the identified
SBC in the \He\ and \CIII, \HeI, \Hb\ lines,
respectively, from the BTA and Subaru observations
in 2007. It should be noted that the spectral resolutions
at these telescopes are different, $R\approx 1000$ and $\approx 2500$, respectively.

Figure~\ref{fig:HeII_07} confirms that the SBCs of the \He\
line and the \CIII\ blend cannot be measured
separately, but the combined SBC is recorded with
confidence. In particular, it can be clearly seen that
the SBCs of these lines are strongly eclipsed by the
donor star. The \HeI\ $\lambda4713$ emission line and a diffuse
interstellar band (DIB, $\lambda 4726.3$) are seen in the same
figure rightward of the \He\ line; both these features
are on the red wing of the SBC in the \He\ line.

The red asymmetry of the SBC in the \HeI\ line is
clearly seen in Figure~\ref{fig:HeI_07} . This line may be considered as
a single one, although it blends with the Fe\,II~$\lambda 4923$
line; the iron lines are considerably weaker than the
helium lines in the spectrum of SS~433. In addition,
both lines are very close in position. The SBC of the
\HeI\ line is significantly eclipsed by the donor star,
just as in the \He\ line.

The SBC of \Hb\ is shown in Fig.~\ref{fig:Hb_07}. It is narrower
than the corresponding components of the  \He\ and
\HeI\ lines; its profile is symmetric. Even if there is
an asymmetry in the SBC of \Hb, it is more likely
the opposite one: the blue wing of this SBC may be
brighter. The SBC in \Hb\ shows no orbital eclipses.


\begin{figure*}[t]
\centering
\includegraphics[width=0.45\textwidth]{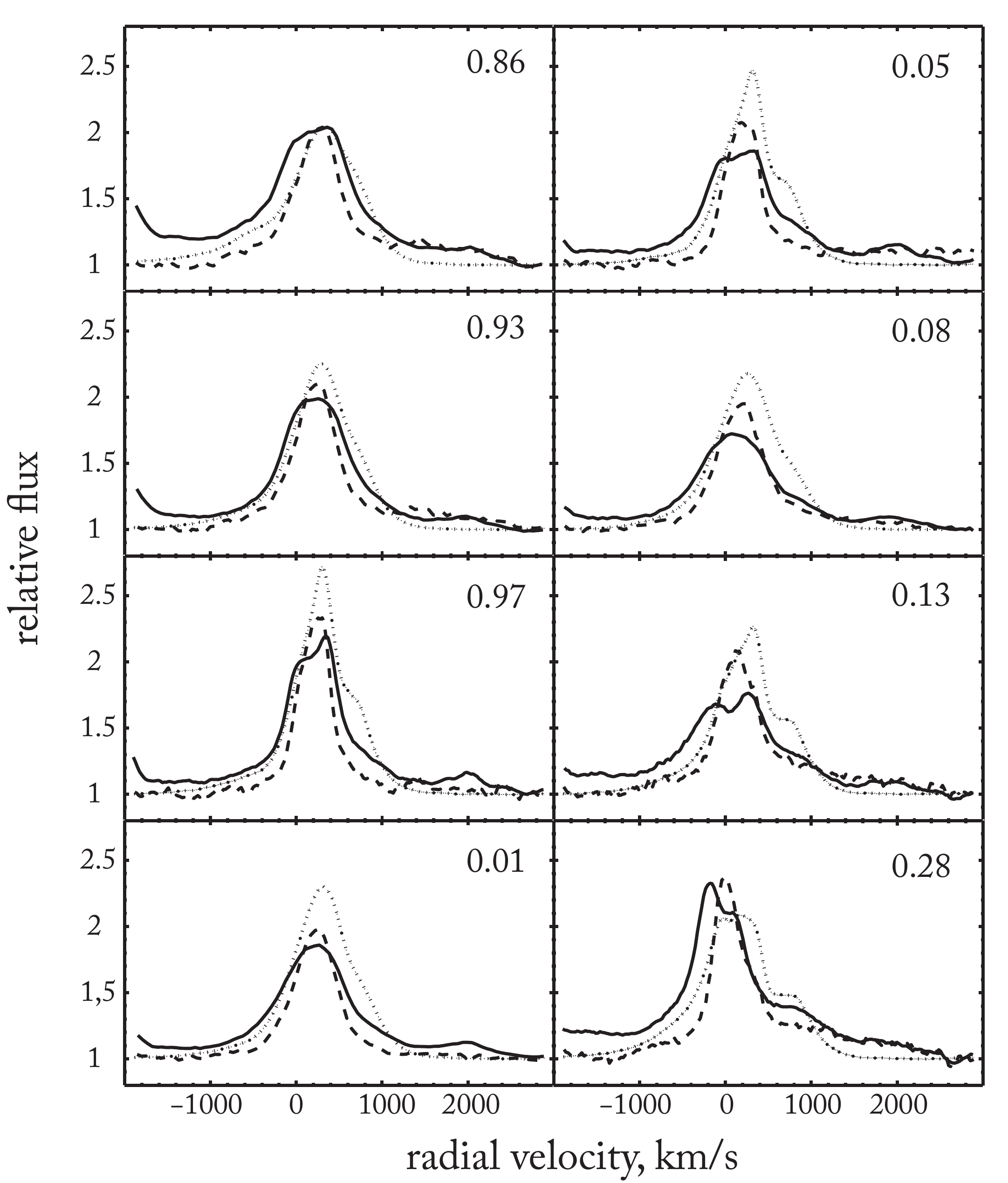}
\caption{Comparison of the \He\ (thick solid lines), \HeI\ (dashed lines), and \Hb\ (dotted lines) profiles with SBCs from the
2007 data. Here, we introduced such a coefficient for the \Hb\ and \HeI\ lines that the intensities of these lines at the first date of
observations were equal to those of the \He\ line. We see eclipses in \He\ and \HeI\ and no eclipses in \Hb. Outside eclipses,
the red wings of the SBCs in the \He\ and \HeI\ lines are identical.
}
\label{fig:HeI_04}
\end{figure*}
Figure~\ref{fig:HeI_04} compares the \He, \HeI\, and \Hb\ line
profiles on the scale of radial velocities from the same
spectra for 2007. In this figure, we introduced such
a coefficient for the \Hb\ and \HeI\ lines that the intensities
of these lines at the first date of observations
(orbital phase $\phi = 0{.}855$) were equal to those
of the \He\ line. This allows the relative eclipses in
these three lines, when the region surrounding the
relativistic object is eclipsed by the donor, to be seen.
Comparison of the relative intensities shows that the
\Hb\ formation region is not eclipsed at all, while the
\He\ formation region is eclipsed deeper than the \HeI\
one.

When comparing the profiles in Figs.~\ref{fig:HeI_07}--\ref{fig:HeI_04}, it
should be kept in mind that the spectra were taken
with different spectral resolutions. The total line
flux does not depend on the resolution, while the
line height does. The eclipse and out-of-eclipse
profiles obtained with the same instrument should be
compared.

It follows from Fig.~\ref{fig:HeI_04}  that the red wings of the
SBCs in the \He\ and \HeI\ lines are virtually identical
in the out-of-eclipse spectra, although they differ
during eclipses. The blue wing of the \He\ line is
distorted by the contribution from the \CIII\
blend. The SBCs of the \He\ and \HeI\ lines extend to
velocities of $\sim 2500$\,km $\mbox{s}^{-1}$. The red wing of \He\
exhibits the \HeI\ $\lambda 4713$ line; when recalculated to
the radial velocities of  \He, this line is located at
1730~km $\mbox{s}^{-1}$. On the contrary, the red wing of \Hb\ has a different shape and evolves differently with orbital
phase.

The identity of the red wings of the SBCs in the
\He\ and \HeI\ lines, especially outside eclipses when
the SBC formation and visibility regions are not distorted
by the donor's partial eclipse, suggests that
these components in the \He\ and \HeI\ lines are the
same in nature. The SBC profile in the \He\ line
(the blue wing blends with \CIII) may also be
asymmetric, just as the SBC profile in the \HeI\ line.
The SBC of \Hb\ is different in origin.

The individual components in the line profiles and
their behavior with orbital phase are clearly seen in the
Subaru spectra in Fig.~\ref{fig:HeI_04}. Two narrow components in
the \He\ and \Hb\ lines and only one narrow component
in the \HeI\ line are clearly identified. A broad
red component is prominent in all three lines, with its
relative intensities being different in different lines.

\subsection{Orbital and Precessional Variability}
\label{sec_var}

\begin{figure*}
\centering
\includegraphics[width=0.6\textwidth]{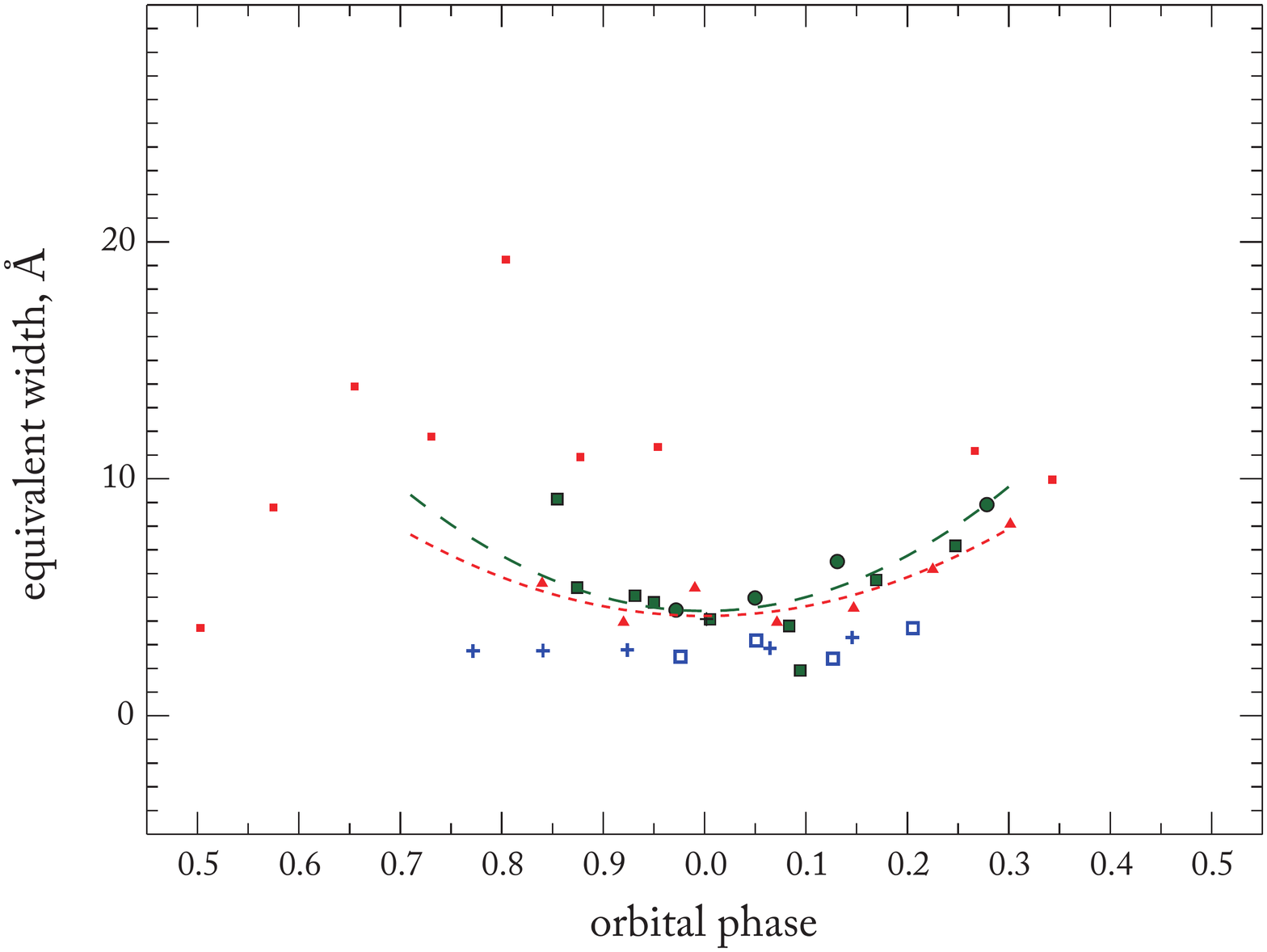}
\caption{Equivalent width of the sum of the \He\ and \CIII\ SBCs versus orbital phase. The squares, triangles, crosses,
and circles correspond to the BTA, Gemini, WHT, and Subaru observations, respectively. The filled symbols correspond to
precessional phases near zero (the disk is maximally open to the observer); the hollow blue squares and blue crosses correspond to the
crossover times. The small red squares and red triangles represent the active state. The lines with green long and red short dashes indicate the
course of the eclipse, respectively, in the passive and active states of SS~433 for the maximally open disk (only the Gemini
data).}
\label{fig:EW_HeII}
\end{figure*}

In Fig.~\ref{fig:EW_HeII}, the equivalent width of the sum of the
\He\ and \CIII\ SBCs is plotted against the
orbital phase. A clear SBC eclipse is observed at the
precessional phases of the maximally open disk ($\psi\approx 0$, the filled symbols in the figure). The observations
at times close to the crossover (hollow squares and
crosses) are also presented there; no SBC eclipses are
observed at these phases, with the equivalent width
of this component being small, 3--4\,\AA. The eclipse
of the SBC formation region by the opaque conical
wind (photosphere) of the accretion disk of SS~433
at edge-on disk phases can be a possible interpretation.
In this case, the SBC of the \He\ line can be
formed in the same place where the optical emission
from the supercritical accretion disk originates. As
is well known, during the crossover, the amplitude
of the eclipses by the donor is small, their shape is
irregular, and the system's total optical brightness is
considerably lower than that at a precessional phase
$\psi \approx 0$ \citep{panferov97,goranskij98,cherep02}. Note that even the depths of
the optical eclipses and the SBC eclipses (Table~\ref{tab:2}) are
approximately identical. However, since the SBC intensities
were obtained through a Gaussian analysis,
they may contain a systematic error, in contrast to the
optical brightness that is measured directly.

During the October 7, 2007 observations at the
open-disk phase, the SBC turned out to be weak ($\phi = 0.083$), the same as at the crossover phases. There
was also a weak SBC during the 2003 observations
(Table~1). We did not plot the data for this year
in Fig.~\ref{fig:EW_HeII} in order not to overload the figure. The
SBC equivalent width in 2003 was 3--4\,\AA\ and barely
changed with orbital phase. SS~433 may have been in
a very low state of activity at this time.

In an active state, the SBC in Fig.~\ref{fig:EW_HeII} varies
sporadically, with only the BTA data exhibiting a large
scatter. The Gemini spectroscopy, which began a
week after the completion of the BTA observations,
already shows a regular behavior of the SBC. In active
states, which last from 30 to 90 days, the mean optical
brightness of the object increases approximately by a
factor of 1.5, and powerful flares with characteristic
time scales of hours to days are observed \citep{fabrika04,fabrika03}. In an
active state, neither the times of flares nor the times
of gas ejections in the jets and the wind in \He\ can
be predicted; the active states of SS~433 themselves
cannot be predicted either. During flares, the SBCs
of the \He\ and \CIII\ lines become brighter by
a factor of 2--3. Despite the active state, the Gemini
2006 data, on the whole, fit into the overall picture of
observations in a quiescent state.

\begin{figure*}
\centering
\includegraphics[width=0.65\textwidth]{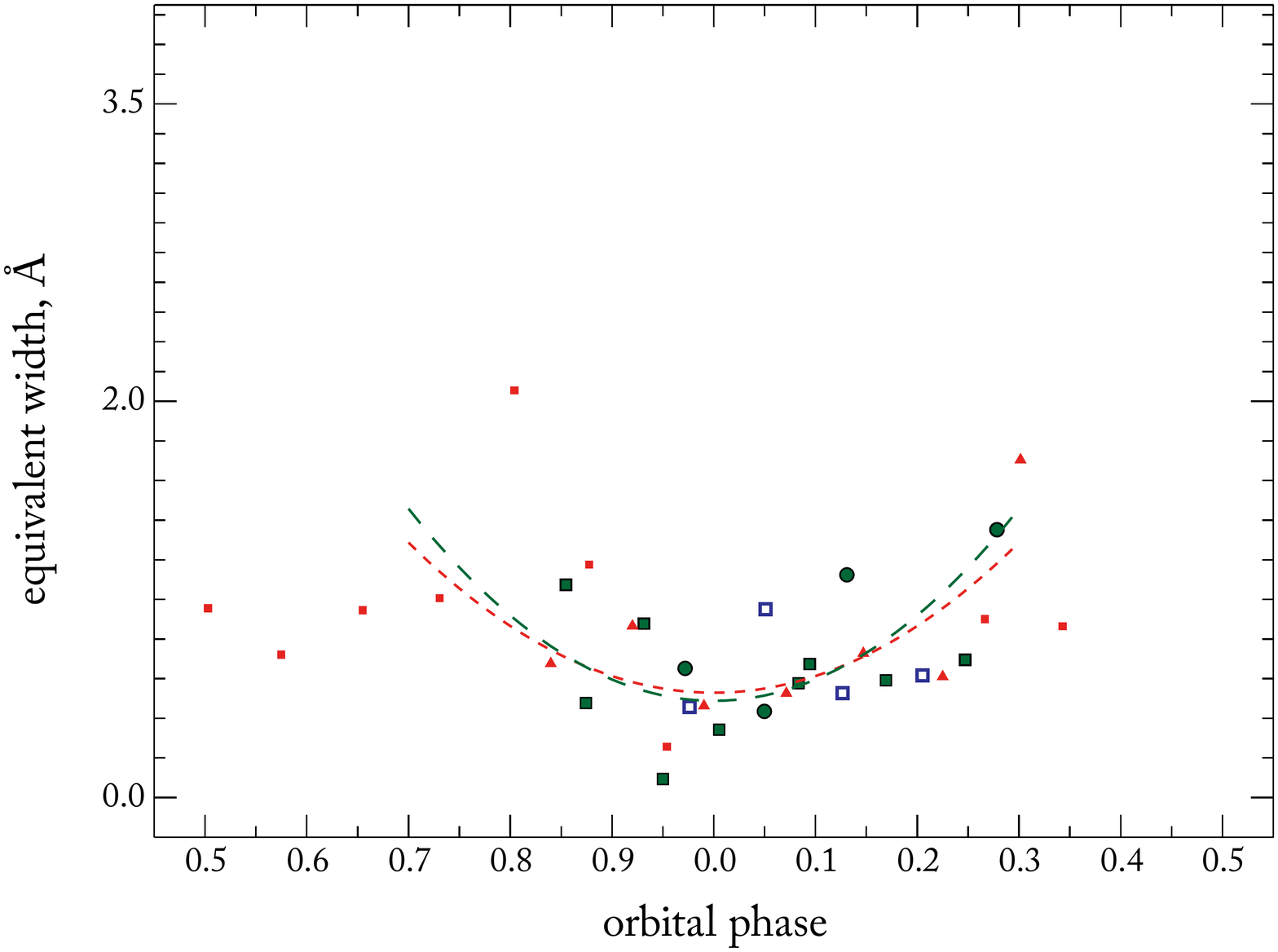}
\caption{Equivalent widths of the SBC in the HeI\ $\lambda 4922$ \AA\ line (without WHT data). The rest is the same as in Fig. ~\ref{fig:EW_HeII}.}
\label{fig:EW_HeI}
\end{figure*}
\begin{figure*}
\centering
\includegraphics[width=0.65\textwidth]{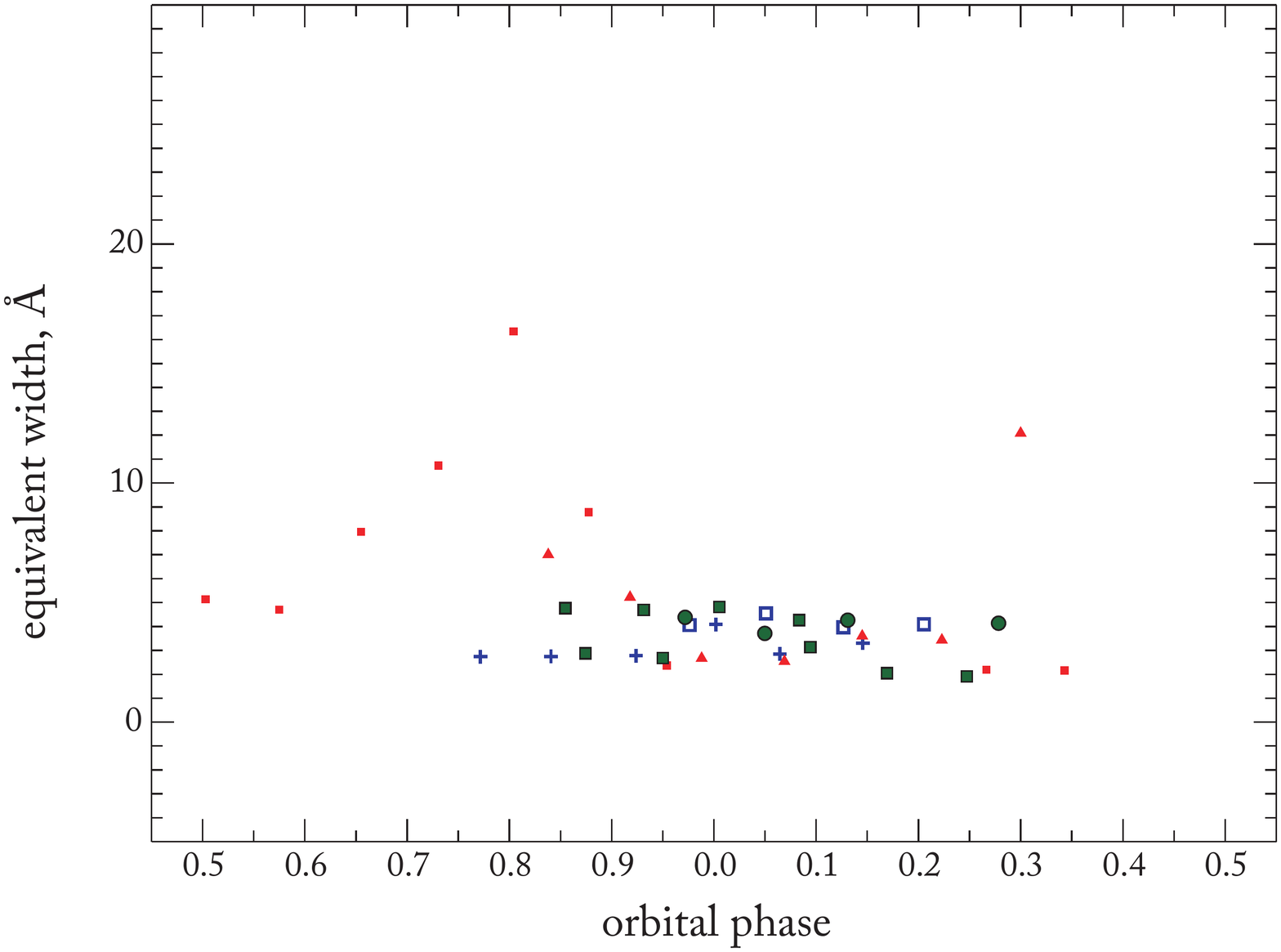}
\caption{Equivalent widths of the SBC in \Hb. The rest is the same as in Fig.~\ref{fig:EW_HeII}.}
\label{fig:EW_Hb}
\end{figure*}
 
\begin{figure*}[t]
\centering
\includegraphics[width=0.6\textwidth]{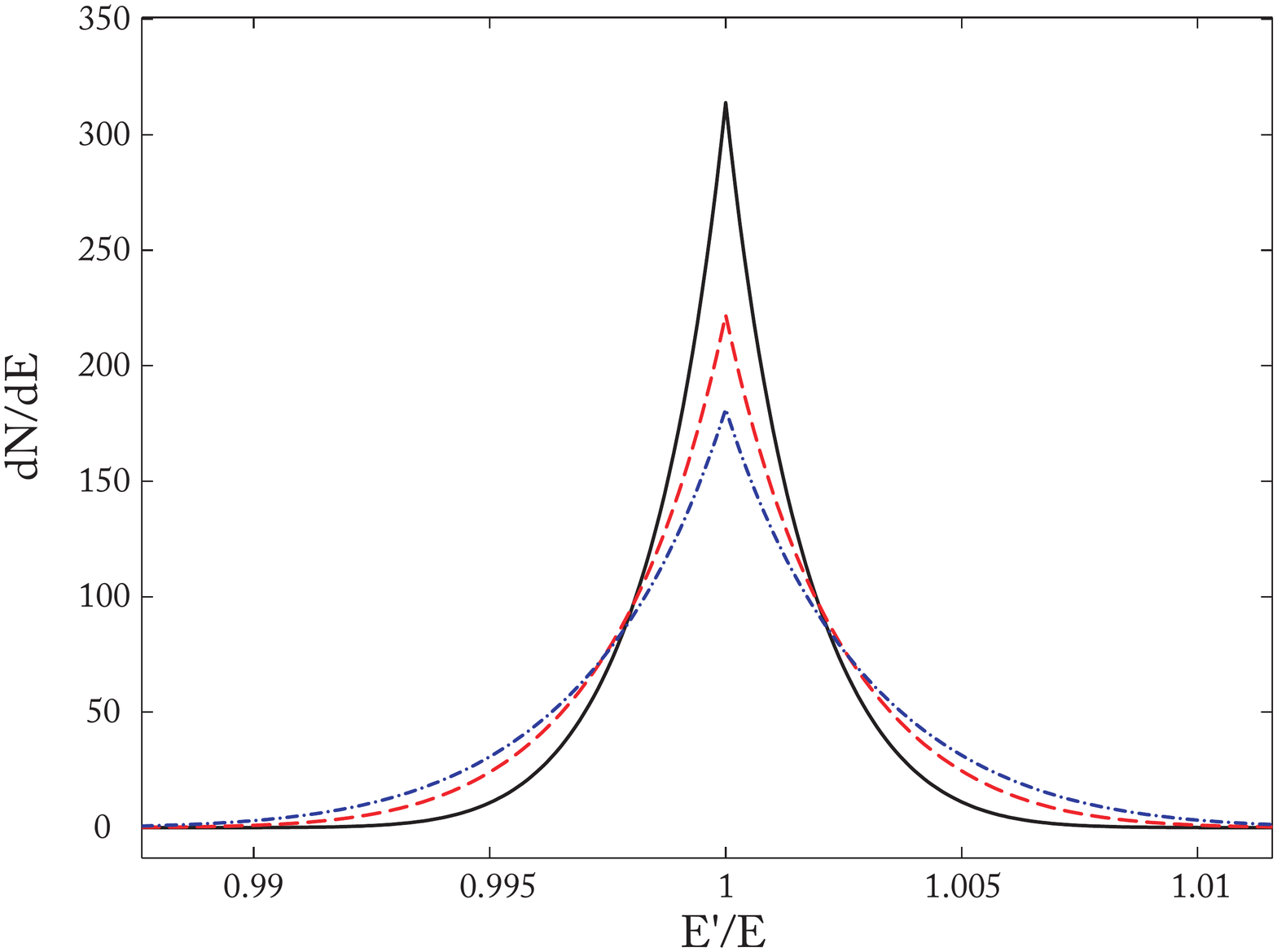}
\caption{Result of single scatterings of the monochromatic \Hb\ line for various electron temperatures of the gas: 1, 2, and 3~eV (black solid, red dashed, and blue dash-dotted lines). $\Delta E  = 0.005$ corresponds to $\Delta \lambda \approx 24$~\AA\ for $\lambda = 4861$ \AA.}
\label{scat_prof}
\end{figure*}

Figure~\ref{fig:EW_HeI} shows the behavior of the SBC equivalent
widths in the \HeI\ line. This line did not fall into
the spectral range of the WHT observations. The
behavior of the SBC in the \HeI\ line agrees well with
the behavior of the SBCs in the \He~$+$~\CIII\
lines: the same deep eclipse near phase $\phi\approx0$ and the
absence of eclipses at the crossover phases. Figure~\ref{fig:EW_Hb}
shows the behavior of the SBC equivalent widths in
\Hb; it is radically different. The SBC of \Hb\ undergoes
no eclipses; its intensity does not depend on the
orbital motion in the system. In spite of this, it is also
sensitive to flares in an active state, just as in the \He\
and \HeI\ lines. Recall that, in contrast to \HeI, the
SBC of \Hb\ has a symmetric profile.

Table~\ref{tab:2} gives the eclipse depths in quiescent and
active states from the fits shown in Figs. ~\ref{fig:EW_HeII}--\ref{fig:EW_HeI} as
well as the mean equivalent widths and FWHM of
the SBCs in these lines. For the ``eclipse depth'' here, we adopted the difference of the values at phases 0.25
and 0.0. For the active state, we provide only the data
from Gemini for 2006. The errors associated with
the choice of the reference points based on which the
components are fitted by a Gaussian make a major
contribution to the measurement error of the SBC
equivalent width. To estimate the fitting error, we varied
the regions of possible positions of the reference
points. These regions are well determined in the line
profiles. The error was calculated by the Monte Carlo
method; the distribution of parameters of all possible
fits for the reference points in specified regions was
considered. For the BTA,WHT, Subaru, and Gemini
spectra, the errors of the equivalent widths ($1 \sigma$), on
average, are 0.3~\AA ~(0.6~\AA ~for WHT).

\begin{table}
\renewcommand{\arraystretch}{1.2}
\centering
\caption{Mean parameters of the SBCs in the
\He\ + \CIII, \HeI\ $\lambda 4922$ \AA\ and \Hb\ lines at
the precessional phases of the maximally open disk in
quiescent (Q) and active (A) states: the eclipse depths $\Delta$\,(EW), the mean equivalent widths and FWHM }
\label{tab:2}
\medskip
\small
\begin{tabular}{l|c|c|c}
\hline
  & \He\ $+$ \CIII\ & \HeI\ & \Hb\ \\
\hline
$\Delta$(EW) (Q) \% & $45{.}4 \pm 5{.}3$& $49 \pm 9$ & 0 \\
$\Delta$(EW) (A), \% & $38{.}1\pm2{.}3$  & $59 \pm 9$ & 0 \\
EW (Q), \AA          & 6.4               & 0.7  & $3{.}7$ \\
EW (A), \AA          & 5.4               & 0.8  & $5{.}2$\\ 
FWHM (Q), \AA        & $87\pm11^1$ & $32{.}1\pm2{.}3$ & $29{.}6\pm1{.}0$ \\
FWHM (A), \AA        & $70\pm4^1$      &$30.0 \pm 3{.}2 $ &$43{.}2\pm 6{.}1$
\end{tabular}
\medskip
  \begin{minipage}{0.95\linewidth}
*The width of the separate \He\ line is $64.0 \pm 8.5$~\AA\ in the
quiescent state and $50.2 \pm 10.8$~\AA ~in the active state.

  \end{minipage}
\end{table}

We found no noticeable changes in the widths of
the SBCs in these lines with orbital or precessional
phase. Table~\ref{tab:2} gives the mean SBC widths for a precessional
phase $\psi\approx0$. In the crossover, the widths
of the \He\ + \CIII\ and \HeI\ SBCs increase
insignificantly, by 7 and 10\,\%, respectively; the width
of the SBC in \Hb\ remains unchanged. It follows
from Table~\ref{tab:2} that during the active state of SS~433,
the SBC of both the Bowen blend and the separate
\He\ line narrows, while the SBC of \Hb\ broadens. In
the active state, the data only from Gemini (2006) are
presented. The width of the sum of the \He\ line and
the \CIII\ blend may not change greatly, because
its value is determined mainly by the separation between these two lines. We also provide the width of
the separate \He\ line. However, it should be kept in
mind that the error in the FWHM of this line from the
blend can be significant.

We propose that the SBCs of the \He\ + \CIII\ and \HeI\ lines are similar in origin, because their
behavior is in many respects similar. These components
can be formed in the high-velocity wind from
the supercritical disk of SS~433. The opening angle
of the wind funnel can be such that the width of
the line SBC will be weakly sensitive to a change
in disk precessional phase. The SBCs of these
lines are formed in the same place where the optical
emission originates: they are eclipsed by the donor.
The absence of eclipses at edge-on disk precessional
phases is also understandable: the outer edge of the
funnel in the wind covers the inner parts of the funnel.
The SBC formation region is located in the funnel of
the wind from the supercritical disk. The asymmetric
profile of the He I SBC can be a P\,Cyg profile in
the wind from the supercritical disk. Below, we will
consider the model for the SBC formation in the \He\
and \HeI\ lines in more detail.

The SBCs in \Hb\ and the helium lines differ in
many respects. The SBC of \Hb\ exhibits no eclipses,
and it has a relatively symmetric profile. The SBC
width in \Hb\ is comparable to that in the \HeI\ line,
but the SBC profile in the latter is significantly asymmetric.
Below, we will check whether the SBC of \Hb\
can be formed through the scattering of radiation by
free electrons.

\begin{figure*}[t]
\centering
\includegraphics[width=0.6\textwidth]{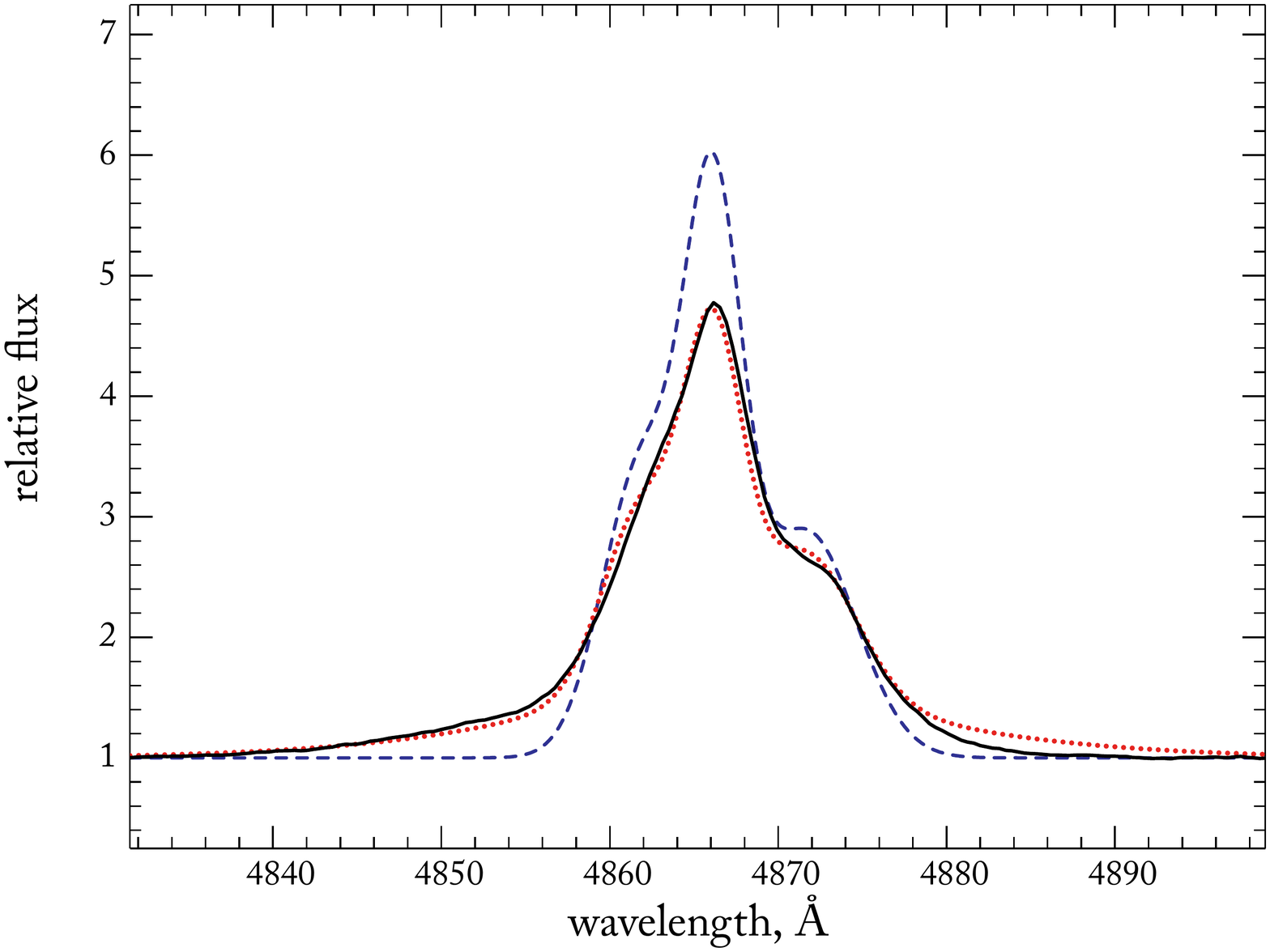}
\caption{Comparison of the observed \Hb\ profile (black solid line) at $\phi =0.97$ with the model profile after scattering (red dotted line) that
was obtained when a fraction $\tau \approx 0.25$ of photons from the initial model profile (blue dashed line) was scattered.}
\label{Hb_model_v2bw}
\end{figure*}


\section{SBC FORMATION IN \Hb: THE SCATTERING MODEL}
\label{sec_Hb}

If the typical width of the SBCs in the \HeI\ and
\He\ lines is taken to be FWHM~$1800$--$2000$\,km $\mbox{s}^{-1}$,
then we will find that for the SBCs to be formed in
the disk, its characteristic size must be $\sim 3 \times 10^{10}\,$ cm,
which is smaller than the size of the donor in SS~433
by tens of times. Such an accretion disk must be
completely covered by the donor in less than half
an hour. On the contrary, fairly long and extended
eclipses of the \He\ SBC formation region are observed.
Because of the powerful outflow, the photospheric
radius of the donor star is greater than the
radius of its critical surface \citep{filippova06}.
Therefore, the SBCs of the helium lines cannot be
formed in the accretion disk. The SBC of \Hb\ cannot
be formed in the disk either, because we observe no
eclipses of this component by the donor star. The
SBCs of our lines cannot be formed in the circumbinary
ring that most likely exists around the system \citep{filippenko88,fabrika93,blundell08,perez10,bowler11} either. The virial velocities of such a ring are an
order of magnitude lower than those we observe in the
SBCs.

The SBC of \Hb\ must be formed in fairly extended
region lest orbital eclipses be observed in it. In addition,
the width of this component does not depend
on precessional nodding in the system. The model of
SBC formation in \Hb\ as a result of scattering looks
most plausible. The mean equivalent width of \Hb\ is
$\approx 40$\,\AA, while the equivalent width of its SBC is \mbox{4--5}\,\AA~(Fig.~\ref{fig:Hb_07} and Table~2). A small optical depth (approximately
equal to the ratio of the equivalent widths of the SBC and the line), $\tau \sim 0.1$, is sufficient for the formation of such a weak SBC.

The line broadening effect arises when photons
from the \Hb\ recombination line are scattered by
free electrons. In this case, the optical depth of the
medium in \Hb\ must be greater than unity. Generally,
the line broadening effect due to electron scattering is
significant at high photon energies (X-ray lines) and
high electron temperatures. However, it is observed in
the optical spectra of luminous blue variables (LBVs)
and hot supergiants \citep[see, e.g.,][]{humphreys11,sholukhova11}. In the case of optical
lines, the relativistic corrections are of no importance,
and the electron scattering is always Thomson one: $h \nu/mc^2 \ll 1$.

The line profile resulting from a single scattering
of monochromatic radiation is determined by the
redistribution function. The redistribution function
is specified as the probability of a photon with frequency
$\nu$ and direction $\Omega$ to have frequency $\nu'$ and
direction $\Omega'$ after its scattering. For the isotropic
problem and the Maxwellian distribution of electrons,
the redistribution function was found with a high
accuracy by \cite{sazonov00}. For the
optical \Hb\ line, it will suffice to use the redistribution
function of the zeroth order of accuracy that describes
the Doppler broadening \citep[see also][]{hummer67}.

Figure~\ref{scat_prof} presents our calculation of single scatterings
of the monochromatic \Hb\ line for various
electron temperatures of the gas (1, 2, and 3~eV). The
profile has broad wings, and the profile wings become
broader with increasing electron temperature. Since
the redistribution function is normalized to unity, the
energy in the line before and after scattering is the
same.

Since the optical depth for scattering is small,
scattering cannot change significantly the shape of
the \Hb\ profile. To model the \Hb\ profile, we chose
the real profile of this line and represented it as three
Gaussian components with FWHM from 200 to
400 km $\mbox{s}^{-1}$ (Fig.~\ref{Hb_model_v2bw}), so that to describe best the
central part of the profile. In this way, we obtained
the model profile before scattering. Subsequently,
we scattered a fraction $\tau$ of photons from this model
profile (in accordance with the optical depth $\tau$ ).
Figure~\ref{Hb_model_v2bw} shows an example for the observation at
an orbital phase of 0.97. Minimizing the residuals
by the least-squares method, we found the probable
domain of parameters: $T_e \approx 20$--$35$~kK and Thomson
optical depth $\tau \approx 0.25$--$0.35$.

It follows from Fig.~\ref{Hb_model_v2bw} that the red wing of the
SBC disagrees with the observed one. In the simple
model of isotropic electron scattering, the line
wings are always symmetric. An asymmetric SBC
profile can be obtained if the region where scattering
occurs moves. In particular, motion with a velocity
from $-250$ to $-300$ km $\mbox{s}^{-1}$ is needed to reconcile the
SBC profiles in Fig.~\ref{Hb_model_v2bw}. As a rule, the emission lines
in the spectrum of SS~433 consist of 3--4 components
(Figs.~\ref{fig:HeII_07}--\ref{fig:Hb_07}) shifted relative to one another by
300--400 km $\mbox{s}^{-1}$. If the parameters $T_e$ and $\tau$ will be
slightly different in different components of the \Hb\
profile, then the asymmetric SBC profiles can also
be explained. Besides, anisotropic scattering could
also be responsible for the asymmetry in the observed
profile. In view of the complexity of the gas flows in the
system, we cannot concretize the scattering model.
Here, it is important that electron scattering may well
explain the SBC in \Hb.

\section{SBC FORMATION IN THE \He\ AND \HeI\ LINES: THE WIND MODEL}
\label{sec_He}

In contrast to \Hb, the SBCs of the \He\ +
\CIII\ and \HeI lines show distinct eclipses by the
donor star at the precessional phases of the maximally
open disk and eclipses by the outer edge of the disk (wind)
at edge-on disk precessional phases. As has been
mentioned above, this means that the helium line
SBC formation region is located in the same place
where the bright optical continuum source is, in the
wind funnel of the supercritical disk.

Above, we have concluded that the SBCs of \He\
and \HeI are similar in nature. The red wings of the
SBCs in these lines are identical, while the blue wing
of the first line is distorted by the Bowen blend; in the
blue wing of the second line, the self-absorption is
observed (which also argues for the Doppler broadening
of the SBCs in these lines). We propose that
the SBCs of the \He\ and \HeI\ lines are formed in
the high-velocity regions of the wind from the disk
of SS~433. Below, we calculate the \He\ line profile
within the model of a wind from a supercritical
accretion disk. Our goal is to show that both the
emergence of a SBC with a width up to $\sim 60$~\AA\ and
the formation of relatively narrow components in the
profile with a width of several hundred km $\mbox{s}^{-1}$ are
possible in this model.

As a result of the high accretion rate in SS~433,
\mbox{$\dot{M_0} \sim 10^{-4}$}\,\solarmass\ $\mbox{yr}^{-1}$, much of the infalling material
is ejected from the system through radiation pressure
in the inner parts of the supercritical disk. Our
model is based on the concept of supercritical disks
by \cite{shakura73}, according to which
the accretion rate in the disk within the spherization
radius $R_{sp} = GM\dot{M}/{2L_{Edd}} \sim 6\times10^9$\,cm is $\dot{M(R)} \sim \dot{M}_0 {R}/R_{sp}$, where $M$ is the mass of the compact
object, and $L_{Edd}$ is the Eddington luminosity. Allowance
for the advection of radiation in the disk within
the spherization radius \citep{lipunova99,poutanen07} does not change the outflow pattern fundamentally.

The velocity of the wind outflowing within the
spherization radius is approximately equal to the virial
one at this radius. Part of the outflowing gas from the
innermost regions is collimated and observed in the
form of relativistic jets from SS~433. The brightest
hydrogen and helium emission lines are formed in
the relatively slow wind (a few thousand km $\mbox{s}^{-1}$).
The photosphere of this wind has a characteristic size ${R_{ph}} \sim 10^{12}$ cm; the temperature of the photosphere is ${T_{ph}} \sim 5 \times 10^{4}$\,K \citep{fabrika04}. If the SBC formation
region is projected onto the hot photosphere (for
example, the far wall of the wind funnel), then a P\,Cyg
profile, i.e., a red asymmetry of the profile, can arise,
which we observe for the SBC of the \HeI\ line.

\begin{figure}
\centering
 \includegraphics[width=1.0\columnwidth]{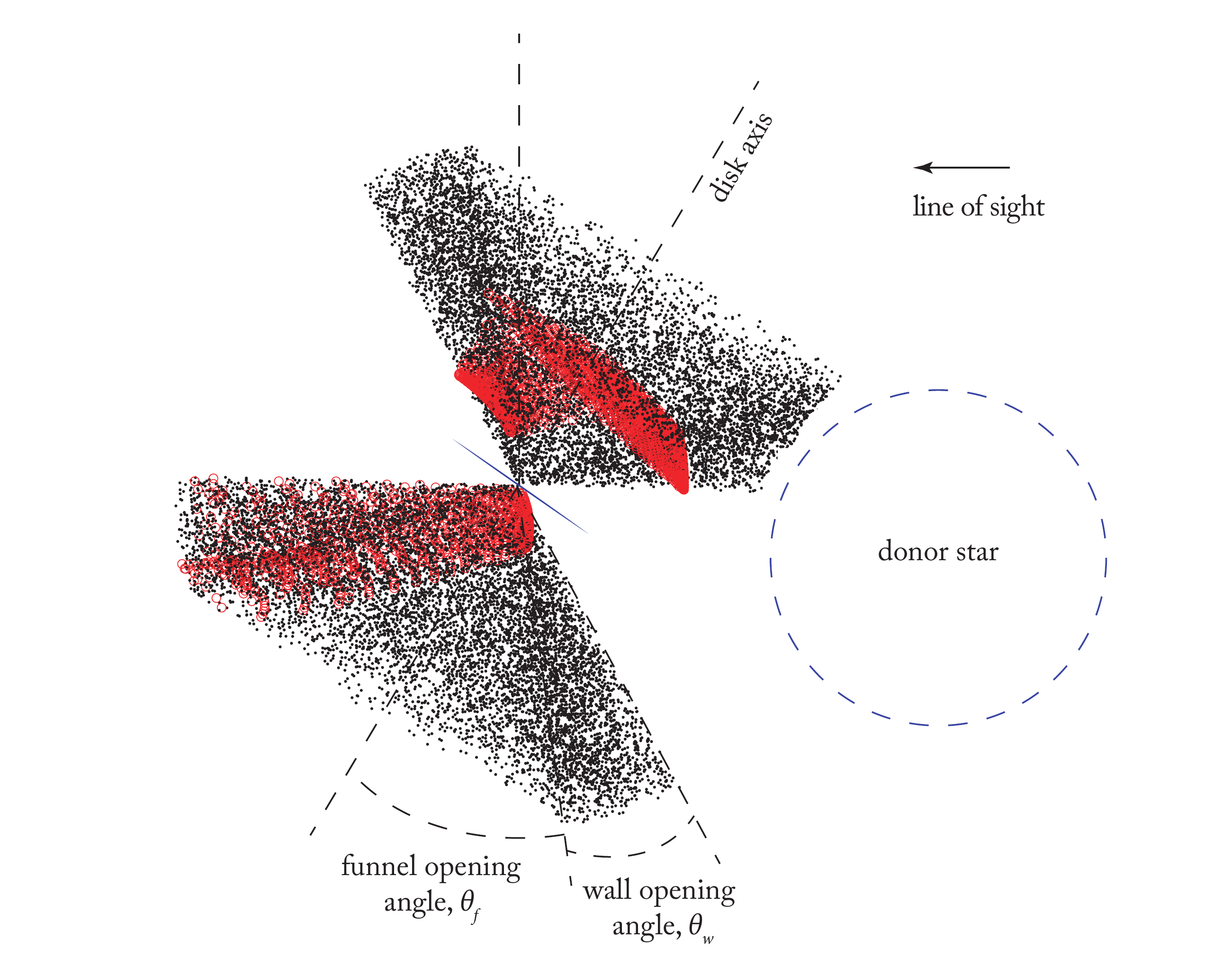}
\caption{Model of the wind from the supercritical disk of
SS~433. The black points indicate the wind. The red
points (which coalesce in the places of their high density)
indicate the wind photosphere as it would be seen by an
observer on the right.}
\label{model}
\end{figure}

We use the \cite{sobolev57} approximation to construct
the model line profile, because the velocity
gradients in the wind exceed considerably the width
of the thermal profile. A simple wind geometry is
specified in the model; the wind is bounded by two
spherical sectors. The wind funnel opening angle
and the funnel wall thickness are model parameters
(Fig.~\ref{model}).

We calculated the wind in a region with a radius
of $4 \times 10^{12}$ cm, which roughly corresponds to the
system's size. In the plane of the sky, the region
was broken down into 100 in {\it y} and 200 in {\it z} (the jet
axis) cells, with the radiation conditions being homogeneous
in each of them. The funnel opening angle
(Fig.~\ref{model}) $\theta_f$  and the wall opening angle $\theta_w$ are model
parameters. Other parameters are the gas ejection
velocity within the spherization radius (the parameter
$\xi$ in the virial velocity $\xi \sqrt{2 G M /R}$), the exponent in
the dependence of the gas emissivity in the line on
radius $(R/R_{sp})^{-s}$, the exponent of the emissivity in
its dependence on polar angle $(\frac{\pi/2 - \theta} {\pi/2})^{p}$ (the funnel
wall heating efficiency), and the radial velocity semiamplitudes
of the compact object and the donor $K_{X}$ and $K_d$. The thin accretion disk (at $R > R_{sp}$) is
also shown in Fig.~\ref{model}, the disk size was calculated \citep{leibowitz84,kallrath99} from the
mass ratio $q = K_d / K_{X}$.

\begin{figure*}[t]
\centering
\includegraphics[width=0.65\textwidth]{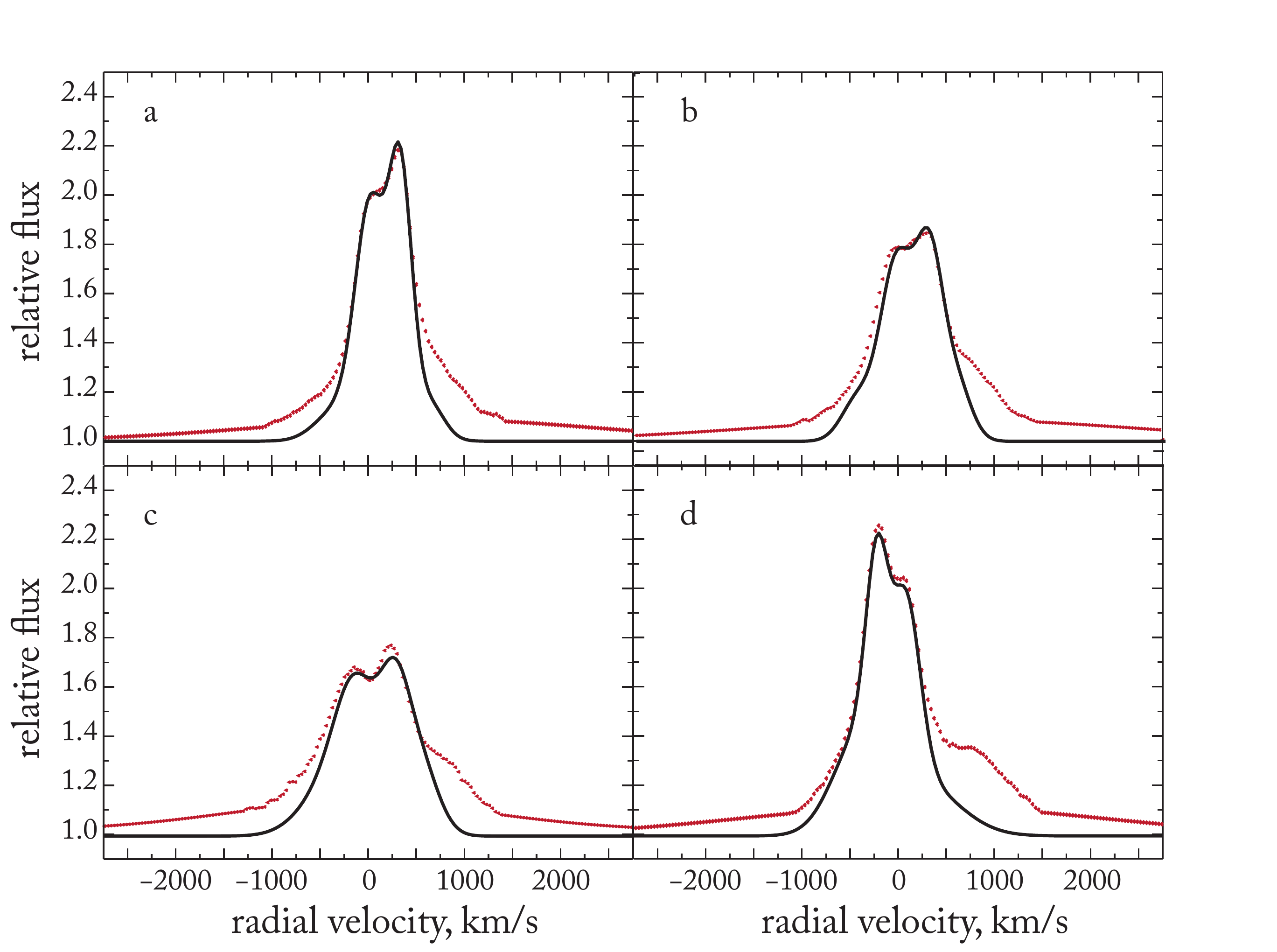}
\caption{Comparison of the model \He\ line profiles (black solid curves) with the observations (red dotted curves) in the model with
mixing. The Subaru observations for 2007; the orbital phases are $\phi = 0.97$ (a), 0.05 (b), 0.13 (c), and 0.28 (d).}
\label{mixmodel}
\end{figure*}

Individual portions of gas were distributed over the
domain of Monte-Carlo computation; we used $\sim10^6$
points in the model. To calculate the line profile, we
summed the radiation from all the wind portions visible
to an observer; the line profile was then convolved
with a thermal profile of width FWHM$ = 20$ km $\mbox{s}^{-1}$.

Individual parts of the wind in the model are
eclipsed by the donor star, the accretion disk, and the
photosphere of the wind itself; the eclipsed regions
are determined by the precessional and orbital phases.
These phases also determine the location of the wind
photosphere that we calculate by integrating the
optical depth for Thomson scattering to $\tau_T =1$ from
infinity along the line of sight. The shape of the
photosphere can be complex. The wind region on
the line of sight located between the photosphere and
observer is assumed to be visible.

We computed two wind models. In the first model
(the model with mixing), the entire wind outflowing
within the spherization radius is averaged over the
momenta, being still deep in the funnel under the
wind photosphere. The wind velocity at this radius is
everywhere the same; the observer sees different radial
velocities due to different projections onto the line of
sight. In the second model (the ballistic model), all
portions of gas move without any interaction along
their trajectories. Having begun its motion from
radius $R$ with an initial velocity $\xi \sqrt{2 G M /R}$, the gas
moves only in the gravity field of the compact object.

The two models are two extreme cases of outward
wind propagation. On the one hand, we fail
to obtain outflow velocities appreciably higher than
$\sim 1000$ km $\mbox{s}^{-1}$ in the model with mixing and with the
known accretion rate in SS~433; on the other hand,
it is obvious that there must be an interaction of the
winds within the spherization radius in the supercritical
disk funnel. The gas outflow pattern can be
very complex. Nevertheless, these two simple models
explain the main trends in the \He\ line profile. Our
model is not self-consistent, but it may well simulate
the Shakura-Sunyaev supercritical disk.

In Fig.~\ref{mixmodel}, the model \He\ profiles constructed
in the model with mixing are compared with the observations.
The orbital and precessional phases in
the model correspond to the real ones. The SBCs in
the ``observed'' \He\ profiles are the Gaussian fits of
these components that were added to the real \He\
profiles (with the subtracted SBC) from Fig.~\ref{fig:HeII_07}. The
model with mixing can reproduce neither the broad
components nor the SBCs. However, it reproduces
well the narrow components originating in the wind.
The pattern of asymmetry in the profile of this line (at
the phases of the maximally open disk before, during,
and after an eclipse by the donor) shown in the figure
is typical of SS~433. Our model does reproduce the typical
evolution of the \He\ profile.

The model parameters with which the observed
profiles are described best (Fig.~\ref{mixmodel}) are the following:
the exponents in the dependences of the gas emissivity
on distance and polar angle are $s = 4$--$5$ and
$p \approx 3$, respectively, the funnel opening angle is $\theta_f \sim 40^\circ$, and the wall opening angle is $\theta_w \sim 20^\circ$.
We estimate these parameters as approximate ones.
Investigating the \He\ profile is not among the goals
of this paper; we are planning to do this in the next
paper. Here, we describe the SBCs of the line profiles.

\begin{figure}
\centering
\includegraphics[width=1.0\columnwidth]{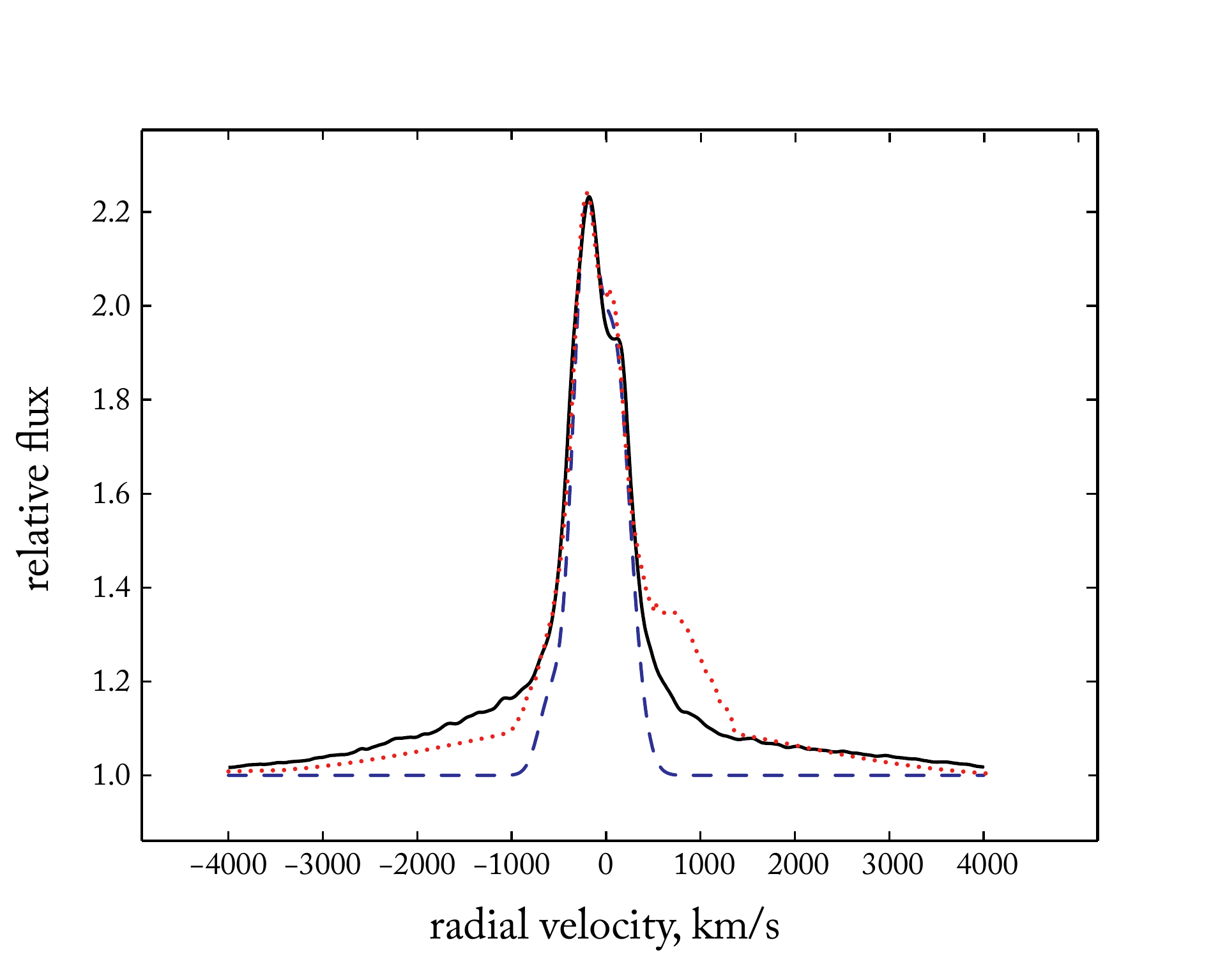}
\caption{Model \He\ profiles. The model with mixing is indicated by the blue dashed line, and the ballistic model is indicated by the black solid line. The red dotted line represents the observed \He\ profile for an orbital phase of 0.28 (October 10, 2007).}
\label{comp}
\end{figure}

Figure~\ref{comp} shows the two model profiles obtained
in the approximation with mixing and in the ballistic
approximation. A SBC appears in the latter case;
consequently, we can explain this component in the
ballistic approximation. Note that we cannot accurately
reconstruct the blue wing of the SBC, because
the \CIII\ blend is in close proximity. None of our
models can explain the broad components (Figs.~\ref{fig:HeII_07}--\ref{fig:Hb_07} and \ref{mixmodel}). In fact, both our approximations are extreme
cases. The interaction of the winds in the supercritical
disk funnel may well give rise to intermediate-width
components.

Above, we put forward the arguments that the
SBCs of the helium lines are formed through a velocity
spread in the wind, and that this component
in \Hb\ is formed through electron scattering. Indeed,
the helium and hydrogen line formation regions are
different; the former are eclipsed by the donor. The
SBC in \HeI\ has a very large red asymmetry that is
difficult to explain by electron scattering. In \He, we
reliably detect only the red wing of the SBC; this component
might be symmetric. All that we know about
the formation geometry of the \He\ and \Hb\ lines is
that they are emitted in the wind from the supercritical
disk, and the formation region of the former line is
considerably smaller than that of the latter one. For
this reason, the optical depth for electron scattering
for the \He\ line must be smaller than that for \Hb;
recall that the optical depth for electron scattering
found by modeling the SBC of \Hb\ is about 0.25--0.35.

An additional factor affecting the role of electron
scattering in the formation of the \Hb\ and \He\ profiles
is the optical depth in these lines. Scattering in a
line increases the effective photon path in the scattering
region, ``tangling'' the photon escape trajectory.

According to \cite{pozdnyakov83}, most of the
photons from a spectral line will undergo at least one
scattering by free electrons if the following condition
is met:
\[
\tau > \left[ 2 \ln\left(\frac{\tau_l}{\tau}\right) \right]^{-1},
\]
Here, $\tau$ is the optical depth for scattering the line photons by free
electrons, and $\tau_l$ is the optical depth in the line. If,
alternatively, the reverse condition is met, then the
line photons, on average, undergo less than one but
more than $\tau$ scatterings by electrons.

The optical depth in a line is defined as $\tau_l = \int n_i  \sigma_i dl$, where $n_i$ is the number density of the ions corresponding to the transition, $\sigma_i$ is the absorption
cross section that for the line center is
\[
\sigma_i =  \frac{e^2 \sqrt{\pi}}{mc} f_i  \frac{1}{\Delta v_D},
\]
here, $f_i$ is the oscillator strength of the transition
under consideration, $\Delta v_D  = v_0 \sqrt{2kT/mc^2}$ is
the Doppler width of the line. For the $4d^1\rightarrow 2p^1$
transition in the hydrogen atom, $f_{H\beta} \approx  1.096 \times 10^{-1} $, while for the $4s^1\rightarrow 3p^1$ transition in \He,
$f_{HeII}  \approx 3.226 \times 10^{-2} $ (the AtomDB database; \cite{foster12}). At the same temperature, we
find $\sigma_{H\beta} / \sigma_{HeII} \approx 1.77 $.

Let us compare the optical depths in the \He\
and \Hb\ lines by assuming their formation regions to
be identical. At a mass loss rate in the wind from
SS~433 $\dot M_w \sim 10^{-4}$\,M$_{\sun}$\, $\mbox{yr}^{-1}$, an outflow velocity of
$\sim 1000$ km $\mbox{s}^{-1}$, and distances of $\sim 10^{12} - 10^{13}$ cm,
the particle number density in the wind can be $3 \times 10^{10-12}$\,cm$^{-3}$.

Estimating the number of particles corresponding
to the transitions in these lines in the Saha-Boltzmann approximation, we find that the condition
$\tau_{H\beta} / \tau_{HeII} > 1$ is met at temperatures $T < 25000$--$30000$~K. Since \Hb\ is emitted in a much
more extended region and at lower temperatures, the
appearance of scattered components in this line is
much more probable than in the \He\ line.

\section{CONCLUSIONS}

We detected new components in the profiles of
main stationary lines for SS~433 (the hydrogen, \HeI,
\He\ lines and the Bowen \CIII\ blend) and
investigated these components in the \Hefn,
\HeIfn, and \Hb\ emission lines. SBCs are
recorded with confidence in the profiles of the brightest
lines. They are low-contrast substrates with a
width of 2000--2500 km $\mbox{s}^{-1}$ in \HeI\ and \Hb\ and
4000--5000 km $\mbox{s}^{-1}$ in \He\ that are considerably
broader than the lines themselves. It is unlikely that
the SBCs will affect the accuracy of measuring the
radial velocities of the lines in SS~433 because of their
large width.

We found that the SBCs of the \He\ and \HeI\
lines are eclipsed by the donor star. The behavior of
the SBC eclipse depth for these lines as a function of
the disk inclination to the line of sight (precessional
phase) turned out to be similar in main features to
the behavior of the optical brightness of SS~433. The
same component in \Hb\ shows no variability with
precessional phase and has no orbital eclipses. This
leads us to conclude that the SBCs of the helium
and hydrogen lines are different in origin. Since the
SBC of the He\,I\,$\lambda 4922$ emission line has a distinctly
asymmetric P\,Cyg profile, electron scattering cannot
be responsible for the appearance of the SBCs in the
helium lines.

In the case of \Hb, the most probable mechanism
for the formation of this component is the line broadening
through electron scattering. We described a
simple model of the \Hb\ formation region that is extended
enough for the eclipses to be unobservable.
We found that the gas in this region with a temperature
\mbox{$T_e \approx 20$--$35$}~kK (rather closer to the low border) and an optical depth for
Thomson scattering $\tau \approx 0.25$--$0.35$ reproduces well
the SBC width and intensity in \Hb.

The SBCs of the helium lines are formed in the
high-velocity gas of SS~433. We used the model
of supercritical disk accretion proposed by \cite{shakura73}, whose main component is the
presence of a powerful wind from the accretion disk
within the spherization radius. We computed a model
of the wind from the supercritical disk that corresponds
in main features to the Shakura-Sunyaev
approximation. The main patterns of the \He\ line
profiles are well reproduced in this model. We can
explain not only the appearance of the SBCs but also
the evolution of the central two-component part of the
profile of this line during its eclipse by the donor star.

Our model may well be used to analyze the profiles
of the lines in SS~433 originating in the supercritical
disk, but this requires spectroscopic observations
with a resolution of at least $\lambda\,/\delta\lambda = 2000$. We leave a
more detailed study of the behavior of the He\,II\,$\lambda 4686$
line using our model for the next paper.

\acknowledgements
We wish to thank S.~Yu.~Sazonov for helpful
discussions. We used archival data from the 4.2-m William Herschel Telescope (WHT) of the Isaac
Newton Group (ING) at the Observatorio del Roque
de los Muchachos. This work was supported by
the Russian Foundation for Basic Research (project
N\,10--02--0046), the Program for Support of Leading
Scientific Schools of Russia (N\,4308.2012.2), and
the Ministry of Education and Science of Russia
(N\,8406, 8416).

\end{document}